\documentclass[]{spie}  

\usepackage[utf8x]{inputenc} 
\usepackage{amsmath,amsfonts,amssymb}
\usepackage{subcaption}
\usepackage{graphicx}
\usepackage[colorlinks=true, allcolors=blue]{hyperref}
\usepackage{algpseudocode}
\usepackage{algorithm}

\title{Optical design concept of the CMB-S4 large-aperture telescopes and cameras}

\author[a]{Patricio A. Gallardo}
\author[a, b, c]{Bradford Benson}
\author[a, b, d, e, f]{John Carlstrom}
\author[g]{Simon R. Dicker}
\author[h]{Nick Emerson}
\author[i]{Jon E. Gudmundsson}
\author[j] {Richard Hills}
\author[g]{Michele Limon}
\author[a,b,c,e,f]{Jeff McMahon}
\author[k,l]{Michael D. Niemack}
\author[m,n]{Johanna M. Nagy}
\author[o]{Stephen Padin}
\author[p]{John Ruhl}
\author[c]{Sara M. Simon}

\affil[a]{Kavli Institute for Cosmological Physics, University of Chicago, Chicago, IL, USA}
\affil[b]{Department of Astronomy and Astrophysics, University of Chicago, Chicago, IL, USA}
\affil[c]{Fermi National Accelerator Laboratory, Batavia, IL, USA}
\affil[d]{High Energy Physics Division, Argonne National Laboratory, Argonne, IL, USA}
\affil[e]{Department of Physics, University of Chicago, Chicago, IL, USA}
\affil[f]{Enrico Fermi Institute, University of Chicago, Chicago, IL, USA}
\affil[g]{Department of Physics and Astronomy, University of Pennsylvania, Philadelphia, PA, USA}
\affil[h]{Steward Observatory, The University of Arizona, Tucson, Arizona, USA}
\affil[i]{The Oskar Klein Centre, Department of Physics, Stockholm University, Stockholm, Sweden}

\affil[j]{University of Cambridge, Cambridge, UK}
\affil[k]{Department of Physics, Cornell University, Ithaca, NY, USA}
\affil[l]{Department of Astronomy, Cornell University, Ithaca, NY, USA}
\affil[m]{Department of Physics,  Washington University in St. Louis, St. Louis, MO, USA}
\affil[n]{McDonnell Center for the Space Sciences, Washington University in St. Louis, St. Louis, MO, USA}
\affil[o]{California Institute of Technology, Pasadena, CA, USA}
\affil[p]{Physics Dept, Case Western Reserve University, Cleveland, OH, USA}
\affil[q]{CMB-S4 collaboration: https://cmb-s4.org/team/}
\author[q]{the CMB-S4 collaboration}

\authorinfo{Further author information: (Send correspondence to P.A.G.)\\P.A.G.: E-mail: pgallardo@uchicago.edu}

\begin{document} 
\maketitle
\begin{abstract}
CMB-S4 -- the next-generation ground-based cosmic microwave background (CMB) experiment - will significantly advance the sensitivity of CMB measurements and improve our understanding of the origin and evolution of the universe. CMB-S4 will deploy large-aperture telescopes fielding hundreds of thousands of detectors at millimeter wavelengths. We present the baseline optical design concept of the large-aperture CMB-S4 telescopes, which consists of two optical configurations: (i) a new off-axis, three-mirror, free-form anastigmatic design and (ii) the existing coma-corrected crossed-Dragone design. We also present an overview of the optical configuration of the  array of silicon optics cameras that will populate the focal plane with 85 diffraction-limited optics tubes covering up to 9 degrees of field of view, up to $1.1 \, \rm mm$ in  wavelength. We describe the computational optimization methods that were put in place to implement the families of designs described here and give a brief update on the current status of the design effort.

\end{abstract}


\section{INTRODUCTION}
\label{sec:intro}
CMB-S4, the next-generation cosmic microwave background (CMB) experiment, will deliver detailed maps of the microwave sky to advance our understanding of the history of our universe from the high energies at the dawn of time to the growth of structure we observe today. The science case of CMB-S4 is broad and it includes\cite{sciencecase, sciencebook}: the search for gravitational waves predicted from inflation, constraints on dark energy, determining the role of light relic particles in the structure and evolution of the universe, tests of gravity at very large scales, measuring the emergence of clusters of galaxies, new observations of transients at millimeter-wavelengths and the exploration of the outer Solar System. To achieve its ambitious scientific goals, CMB-S4 will require a great increase in sensitivity from current generation experiments\cite{technologybook}. This drastic increase in sensitivity and mapping speed will be achieved by increasing the number of high-sensitivity detectors on the sky and the number of telescopes that observe at the same time. CMB-S4 will make use of two of the best equipped sites in the world for millimeter-wave observations of the CMB: the high Atacama Plateau of Chile and the Geographical South Pole. The CMB-S4 concept will exploit features of the two sites, using the stable atmosphere of the South Pole to map a deep patch of sky and using the ability to observe up to 80\% of the sky from the dry Atacama site\cite{whitepaper}.

\begin{figure}[hp]
    \centering
    \begin{subfigure}{.35\textwidth}
        \centering
        \includegraphics[width=\linewidth]{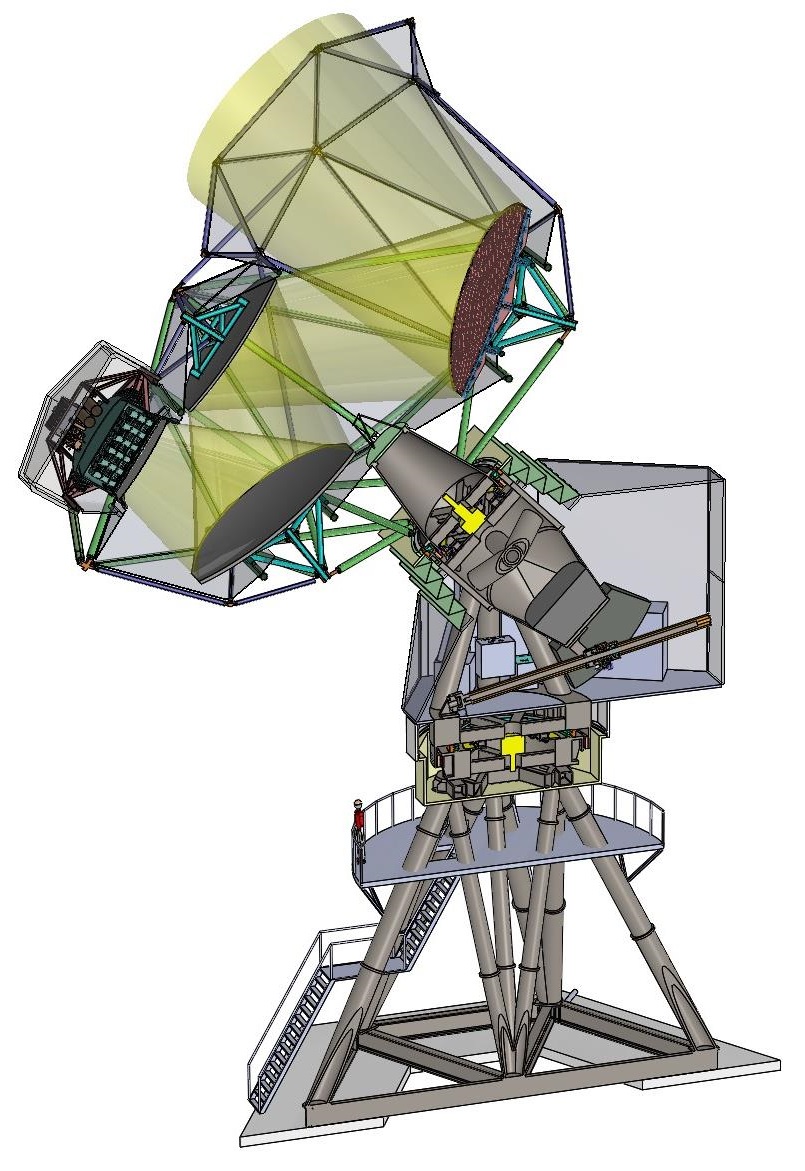}
    \end{subfigure}\begin{subfigure}{.60\textwidth}
      \centering
      \includegraphics[trim={0 0 0 0},clip, width=\linewidth]{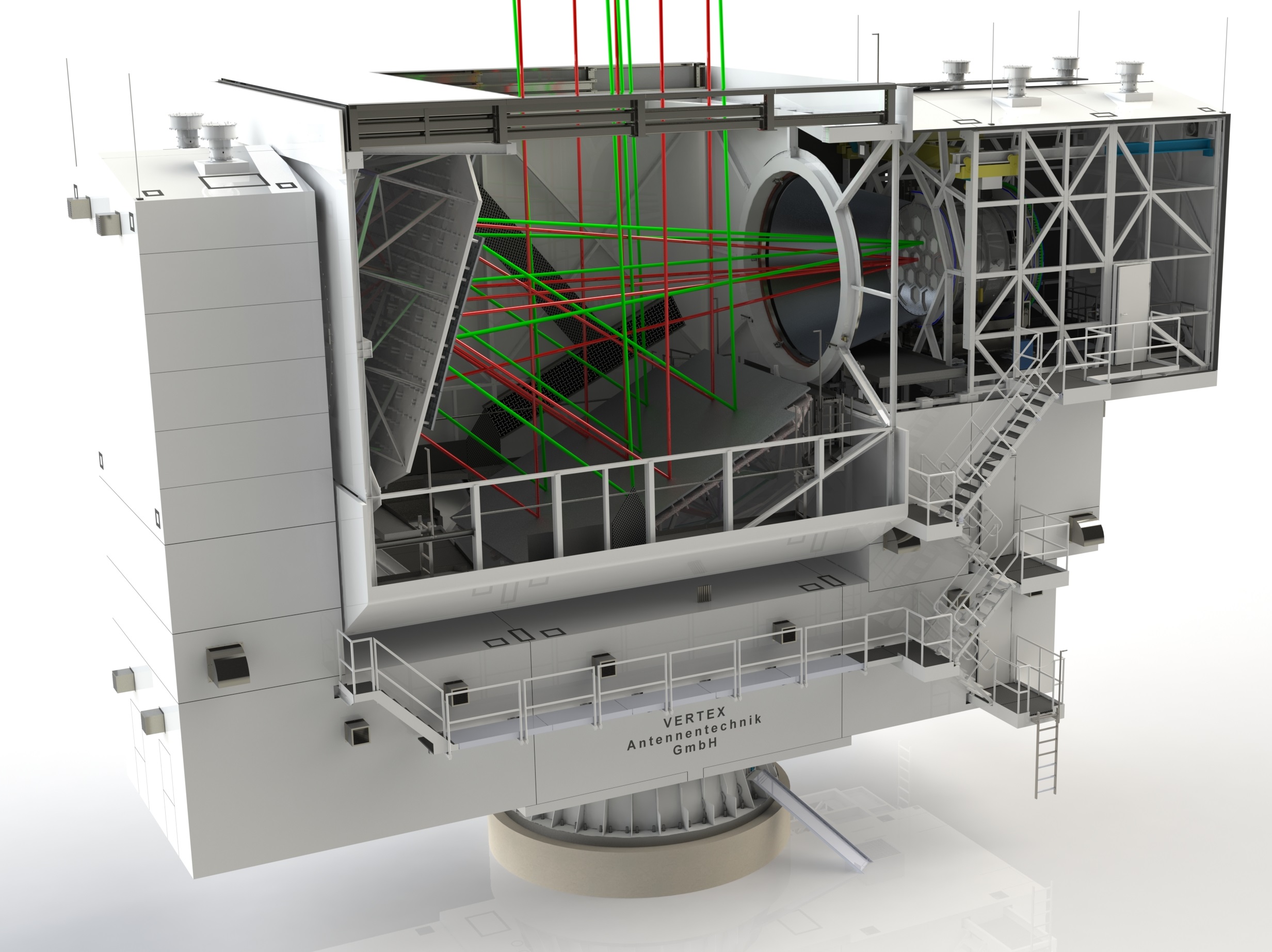}
\end{subfigure}%

    \begin{subfigure}{.4\textwidth}
        \centering
        \includegraphics[width=.8\linewidth]{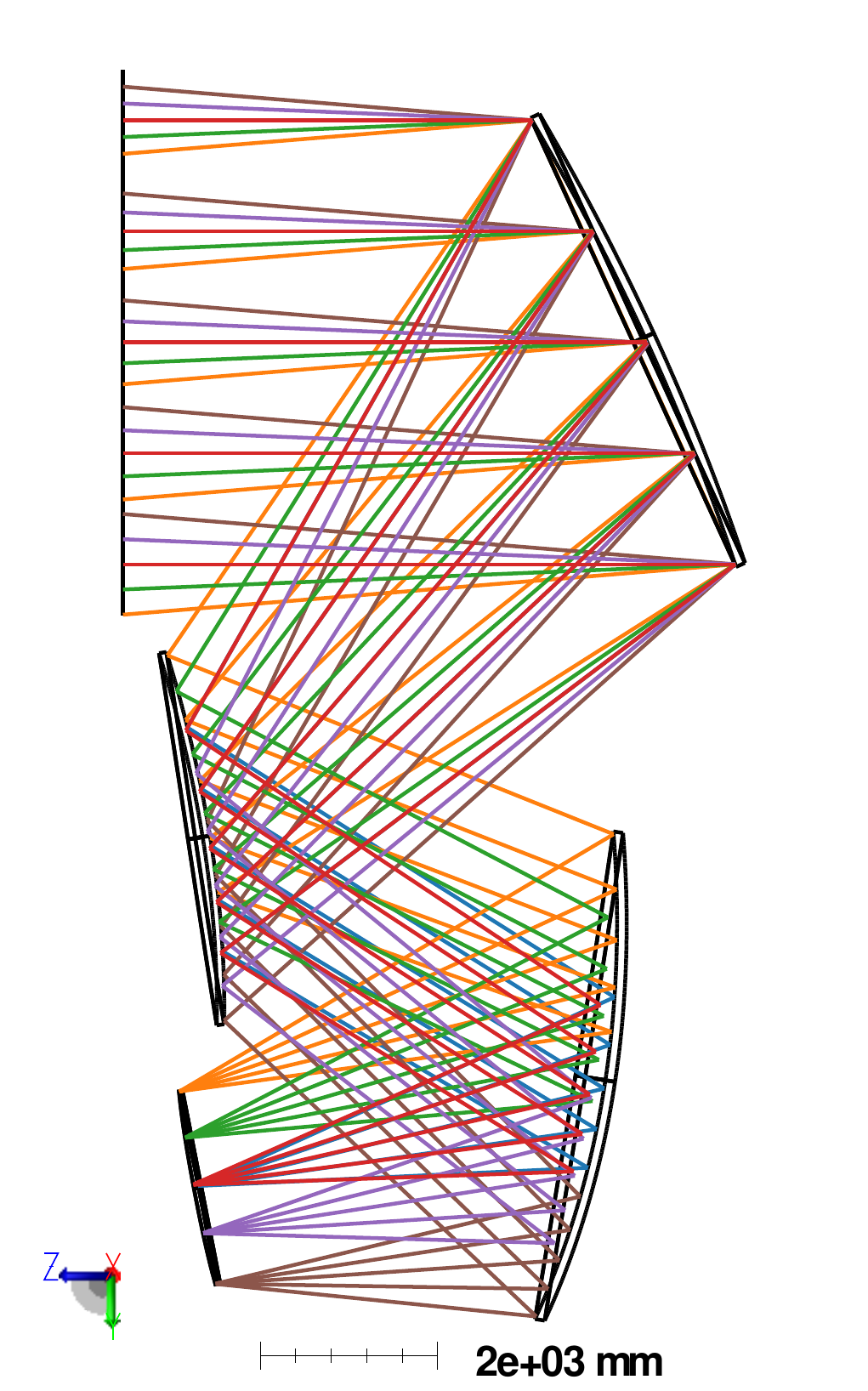}
    \end{subfigure}
  \begin{subfigure}{.5\textwidth}
      \centering
      \includegraphics[width=.8\linewidth]{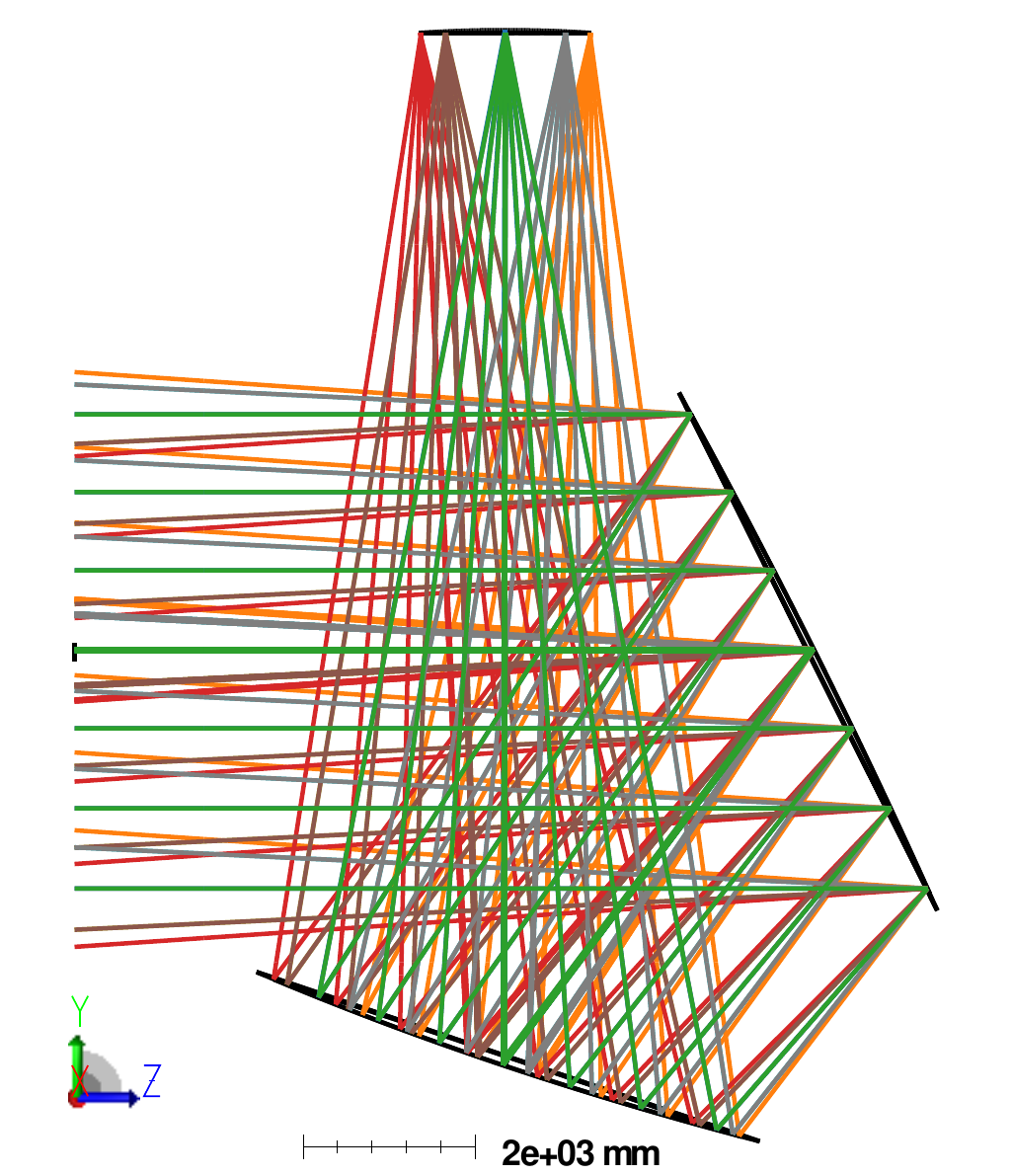}
\end{subfigure}%

    \caption{Top: Rendering of the baseline design concept of CMB-S4. Telescopes are not to scale. Top Left: The South Pole LAT, a five-meter three-mirror anastigmatic telescope. Top Right: The Chilean LAT concept, which is composed of the six-meter paneled crossed-Dragone design\cite{2021ApOpt..60..823G}. Bottom: Optical layout of the two telescope concepts. Bottom Left: Three Mirror Anastigmat optical layout to be used in the SPLAT; this design is an optimized version of the concept presented in Padin 2018 \cite{ApOpt..57.2314P}. Bottom Right: coma-corrected crossed-Dragone optical layout to be used in the CHLAT. This design is identical to the Simons Observatory LAT and the CCAT-prime Fred Young Submillimeter Telescope. \cite{2018SPIE10700E..41P}.}
    \label{fig:renders_layout}
\end{figure}

\begin{figure}[hp]
    \centering
    \begin{subfigure}{0.5\linewidth}
        \centering
        \includegraphics[clip, trim={0.5cm 0cm 1.15cm 0cm}, width=\textwidth]{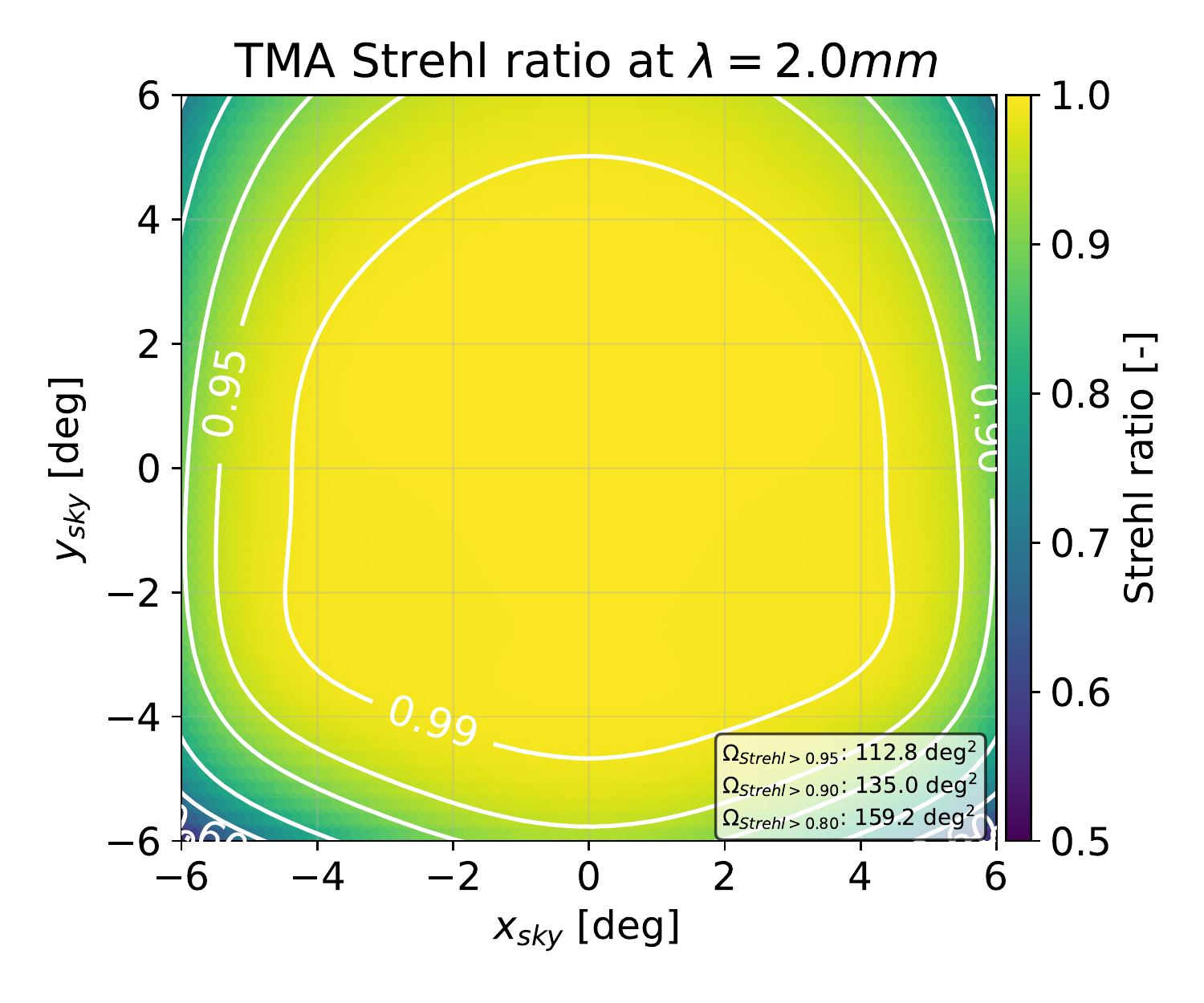}
    \end{subfigure}\begin{subfigure}{.5\textwidth}
      \centering
      \includegraphics[clip, trim={1.15cm 0cm 0.5cm 0cm}, width=\textwidth]{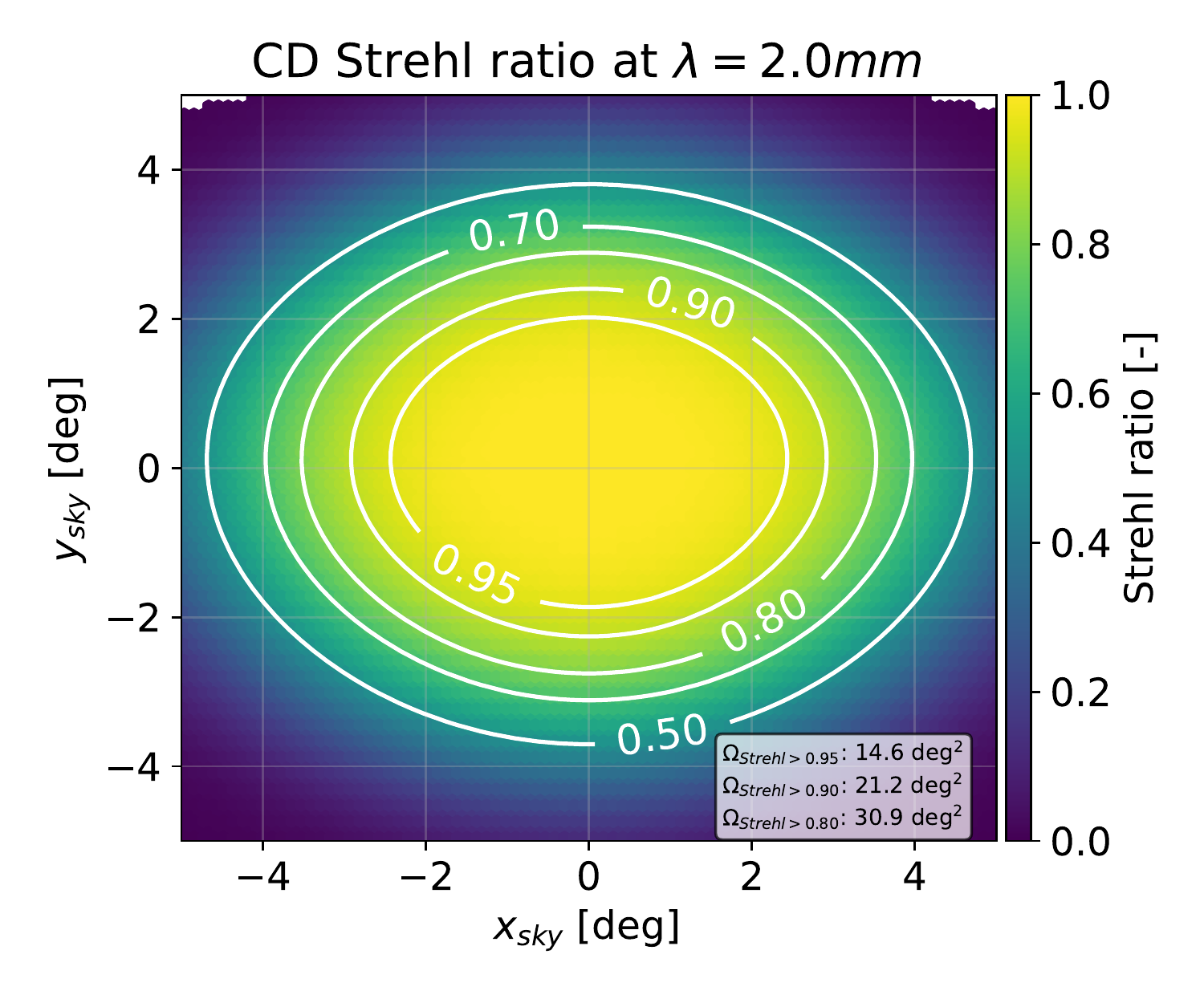}
\end{subfigure}%

    \begin{subfigure}{.5\textwidth}
        \centering
        \includegraphics[clip, trim={0.5cm 0cm 1.15cm 0cm}, width=\linewidth]{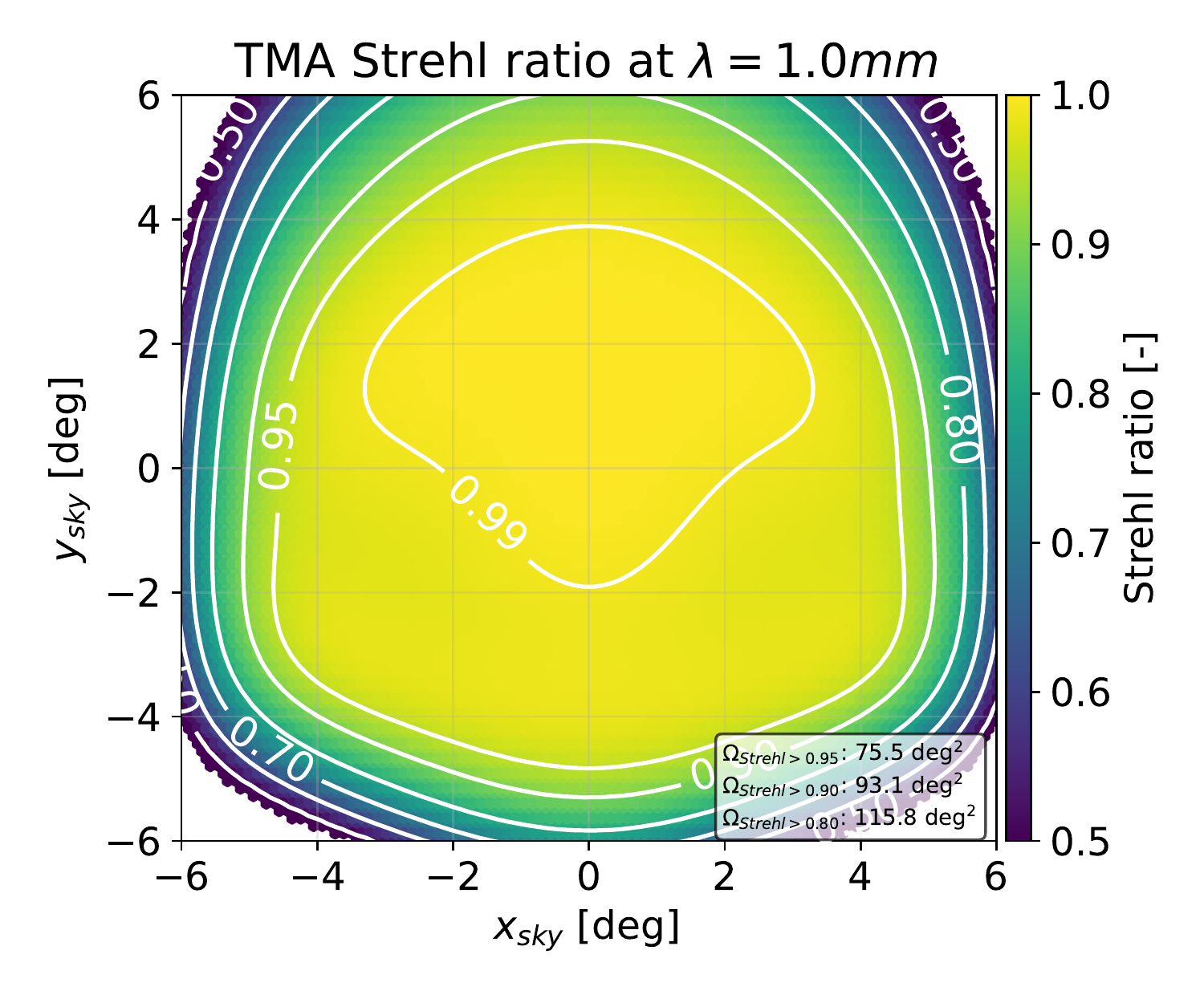}
    \end{subfigure}\begin{subfigure}{.5\textwidth}
      \centering
      \includegraphics[clip, trim={1.15cm 0cm 0.5cm 0cm}, width=\linewidth]{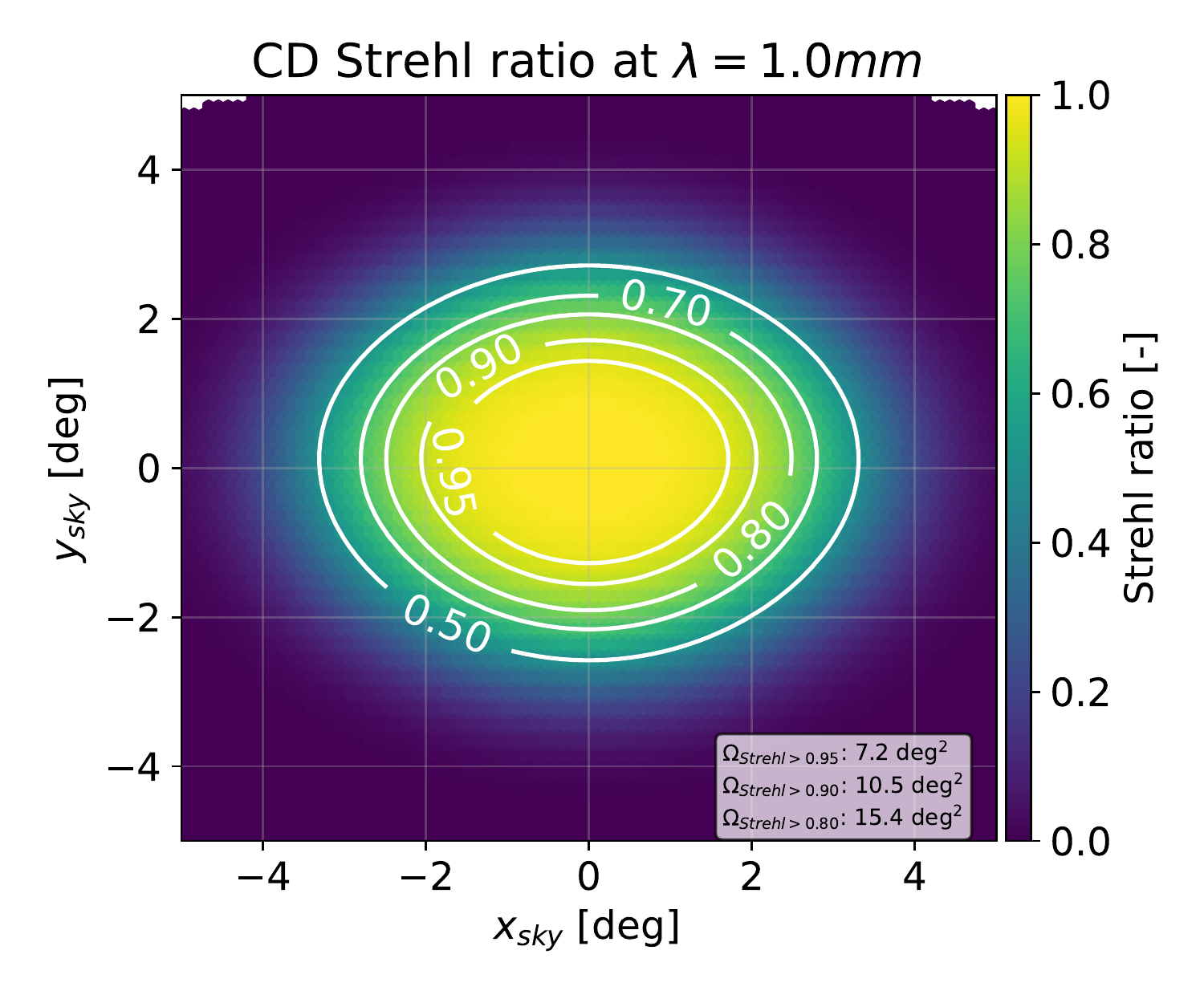}
\end{subfigure}%

    \caption{Strehl ratio in angular coordinates at $2\, \rm mm$ (top) and $1\, \rm mm$ (bottom) wavelengths for the SPLAT TMA (left) and CHLAT CD (right) optical designs. Note that because of the different dynamic ranges in Strehl ratios among the TMA and the CD designs (left vs right), the ranges in the color bars are different (left panel covers 0.5 -- 1.0, while right panel covers 0.0 -- 1.0). The angular extent of the field is wider in the left panels ($\pm 6 \, \rm degrees$ vs $\pm 5 \, \rm degrees$) due to the larger field of view provided by the TMA.}
    \label{fig:LATimgqualcomp}
\end{figure}

In the last decade, large-aperture (six-meter class) CMB telescopes have deployed an increasing number of background limited detectors in order to provide faster mapping speeds at arcminute resolutions\cite{2016JLTP..184..772H, 2014SPIE.9153E..1PB}. This sustained increase in detector count has occurred concurrently with advances in superconducting detector technologies and cold multiplexing. Optical designs of large-aperture telescopes (LATs) have also evolved to accommodate an increasing number of detectors. Currently observing telescopes, which make use of Gregorian designs, can accommodate thousands of transition edge detectors\cite{Fowler:07, Padin:08}. Planned and under-construction CMB observatories, like the Simons Observatory and CCAT-prime, have proposed larger throughput LAT designs, using the crossed-Dragone (CD) optical configuration \cite{2018SPIE10700E..41P}. These observatories are planned to accommodate tens of thousands of detectors. Next-generation telescopes will accommodate hundreds of thousands of high sensitivity detectors to provide a unique view into the early universe\cite{decadal}.

The science goals of CMB-S4 will be met with two surveys: 1) One ultra-deep survey covering 3\% of the sky, observed from the South Pole. This survey will make use of small-aperture telescopes (not discussed in this document). These small-aperture telescopes will be complemented with one five-meter diameter three-mirror anastimatic telescope (TMA). Maps made with observations of the TMA will be used to remove contamination of the degree-scale B-modes caused by gravitational lensing, in a process called delensing. 2) A deep-wide survey covering approximately 60\% of the sky using two six-meter crossed-Dragone LATs located in Chile. The deep-wide survey requires modestly higher angular resolution and its optical design was selected based on the CCAT-prime and Simons Observatory optical designs that are now undergoing construction. The smaller patch observed by the ultra-deep survey requires a uniform distribution of bands across the field of view, so the survey depth is sufficiently uniform. The ultra-deep survey does not require as high angular resolution as the deep-wide survey, and the delensing required to achieve the inflation science goal motivates the pursuit of additional features like the higher and more uniform Strehl ratios provided by the introduction of a third optical surface, the mitigation of systematics provided by gap-less mirrors and boresight rotation.

In this work we present an overview of the baseline optical design concept of the large-aperture telescopes of CMB-S4 and their arrays of 85 imaging cameras. The baseline optical design presented here combines the tested technologies of heritage CMB experiments that are under fabrication with an innovative concept that will deliver higher fields of view and mapping speeds at high frequencies (up to $286\,\rm GHz$, or $1.1\, \rm mm$). We describe the optical concept and the methods used to find optical solutions.  The methods presented here are capable of generating the specific lens prescriptions for a given set of mechanical constraints. The detailed design prescriptions will be presented in a future publication as the opto-mechanical design and the optical design converge. Throughout this document the term diffraction-limited performance is used to refer to Strehl ratios better than 0.8 at a specified wavelength \cite{zmxmanual}.

\begin{figure}[ht]
    \centering  
    \begin{subfigure}{.49\textwidth}
      \centering
      \includegraphics[trim={0.45cm 0 0.49cm 0.93cm},clip, width=\linewidth]{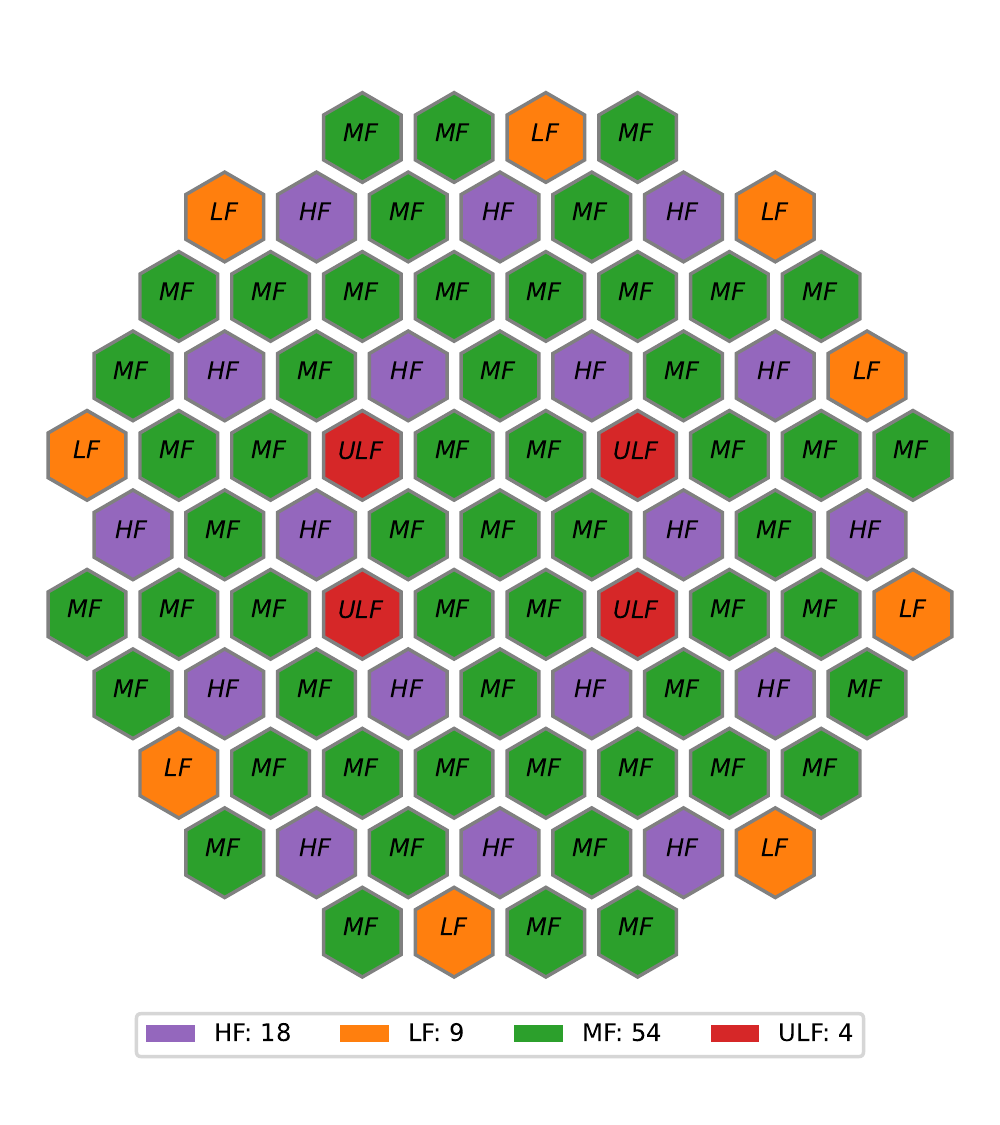}
	\end{subfigure} 
    \begin{subfigure}{.49\textwidth}
        \centering
        \includegraphics[trim={0.45cm 0 0.49cm 0.93cm},clip, width=\linewidth]{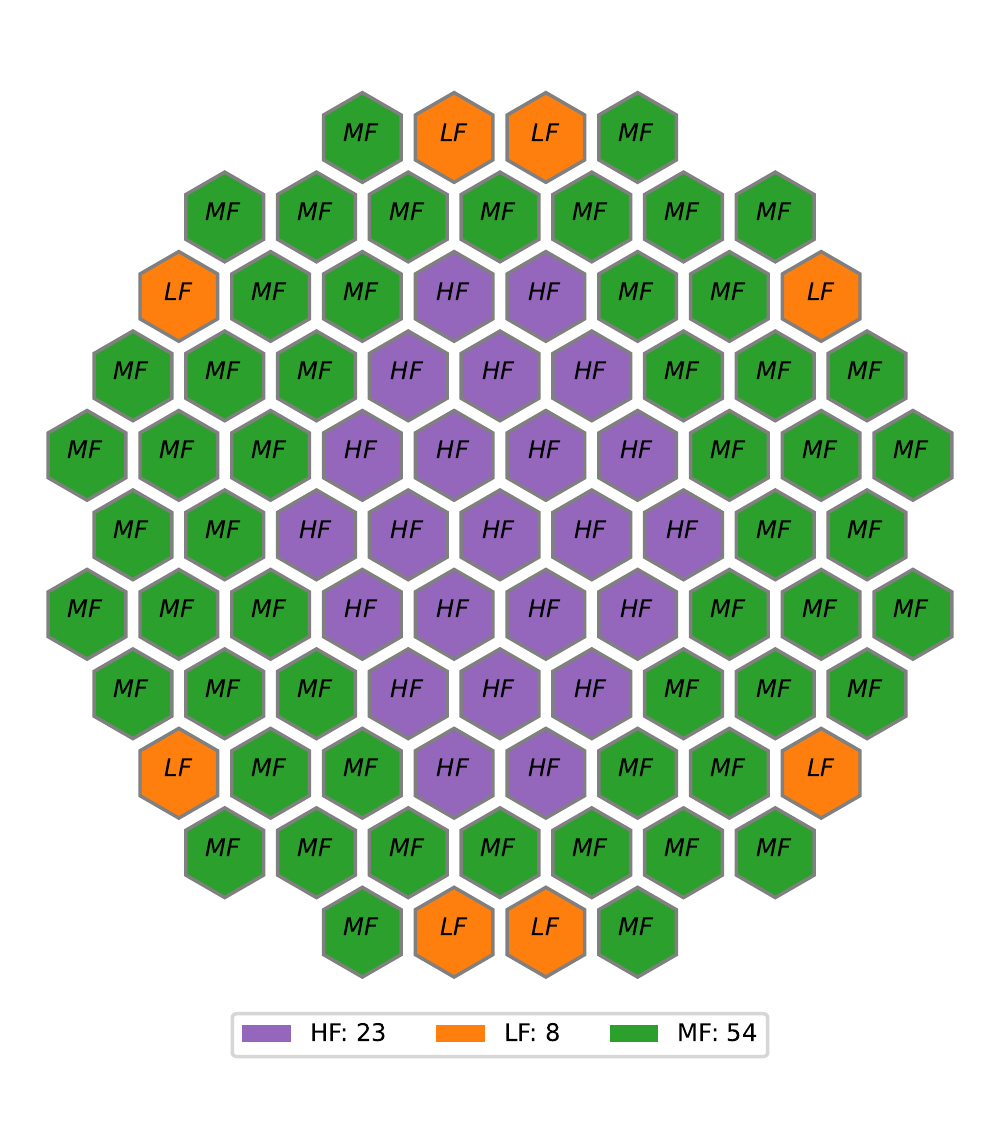}
    \end{subfigure}

    \caption{Camera band distribution for the SPLAT (left) and CHLAT (right). Cameras operating in four  dichroic bands are shown. HF: $286$, $227 \, \rm GHz$ ($1.1$, $1.3 \, \rm mm$, purple), MF: $149$, $92 \, \rm GHz$ ($2.0$, $3.3\, \rm mm$, green), LF:  $39$, $26 \, \rm GHz$ ($7.7$, $11.5 \, \rm mm$, orange) and ULF: $20 \, \rm GHz$ ($15.0\, \rm mm$, red).}
    \label{fig:hexes}
\end{figure}

\section{Design concept}
\label{sec:concept}

The baseline design concept of the large-aperture telescopes for CMB-S4 consists of two types of telescopes: the Chilean Large Aperture Telescope (CHLAT) and the South Pole Large Aperture Telescope (SPLAT). The Chilean site will use a six-meter crossed-Dragone optical design, a two-mirror system that allows a field of view of $6.4 \, \rm degrees$  in diameter at $150\, \rm GHz$ ($2\, \rm mm$ in wavelength).\cite{2018SPIE10700E..41P}\footnote{The crossed-Dragone is capable of $1\, \rm mm$ ($300\, \rm GHz$), or even shorter wavelength, observations albeit with a reduced field of view due to astigmatism on the outer fields\cite{2018SPIE10700E..41P}.} The SPLAT optical design is a five-meter aperture three-mirror telescope, an innovative optical concept that cancels out astigmatism, and therefore offers a wide field of view of 9 degrees in diameter at $286\, \rm GHz$ ($1.1\, \rm mm$ in wavelength), at the expense of increased complexity\cite{ApOpt..57.2314P, 2022Gallardo}. Figure \ref{fig:LATimgqualcomp} shows a comparison of the Strehl ratios across the field of view for these two designs. The reflectors in this three-mirror telescope are gap-less, thus providing low sidelobes when compared to the paneled six-meter crossed-Dragone configuration\cite{2021ApOpt..60..823G}. Figure \ref{fig:renders_layout} (top) shows renderings of these two concepts, while Figure \ref{fig:renders_layout} (bottom) shows their optical layout.


\subsection{CHLAT: crossed-Dragone Design}

The telescopes to be deployed in Chile are arranged in a coma-corrected crossed-Dragone optical configuration. A first physical demonstration of such system is currently under construction for the Simons Observatory and CCAT-prime\cite{2018SPIE10700E..41P}. This optical design is a f/2.6 system, which illuminates a focal plane of 2 meters in diameter, a field of view of 7.8 degrees and can accommodate $\sim $ 100,000 detectors with diffraction-limited performance at 3 mm  wavelength\cite{2018SPIE10700E..41P}. The telescope mirrors are made out of rectangular panels, which are machined to shape and are individually adjusted to achieve phase wavefront coherence. A detailed description of this design concept development and implementation can be found in the literature\cite{10.1117/12.2312985, 10.1117/12.2312971, 2018SPIE10708E..3XO, 2018SPIE10700E..3ED}.

The CD can be used at very high frequencies with a limited the field of view, as will be demonstrated by CCAT-prime. This optical design is optimal for minimal cross polarization.  The gaps between the panels that form the six-meter reflectors produce sidelobes at about $60\, \rm{dB}$ below the main beam at degree angular scales\cite{2021ApOpt..60..823G}. The diffraction-limited field of view of the crossed-Dragone design (at $1\, \rm mm$ and $2\, \rm mm$ in wavelength) is limited by astigmatism, which appears at the highest frequencies in fields far from the boresight. This optical feature can be corrected in the camera system by the inclusion of one biconic lens, as will be discussed later in this document.

\subsection{SPLAT: Three Mirror Telescope}

Three-mirror anastigmatic telescope designs are appealing for millimeter-wave large-aperture systems because they can deliver a wide field of view  that is free of astigmatism down to $1\, \rm mm$ in wavelength. This lack of astigmatism is the result of the additional degree of freedom given by the three surfaces. These three surfaces are able to cancel all three first-order optical aberrations, as opposed to the only two aberrations that can be cancelled using two reflecting  surfaces like in a crossed-Dragone or a Gregorian telescope. An extensive discussion on three-mirror telescope configurations was presented by Korsch (1991) \cite{korsch1991reflective}.

The first proposed three-mirror anastigmat for large-aperture millimeter-wave observations of the CMB was presented in Padin 2018\cite{ApOpt..57.2314P}. This optical design was adapted for CMB-S4 by reducing the f-number to make it similar to the crossed-Dragone's f/2.6\cite{2022Gallardo}. This allows the TMA to have more throughput and share the camera concept with the crossed-Dragone, which has benefits from the opto-mechanical perspective. This redesign also improved image quality at $1\, \rm{mm}$ to excellent values across the whole focal plane. The TMA design consists of a five-meter aperture primary, with a tertiary of a similar size. A detailed description of the TMA for CMB-S4 will be presented in a separate publication (Gallardo et al, in prep\cite{2022Gallardo}).

\begin{figure}[b]
	\centering
	\includegraphics[width=0.8\textwidth]{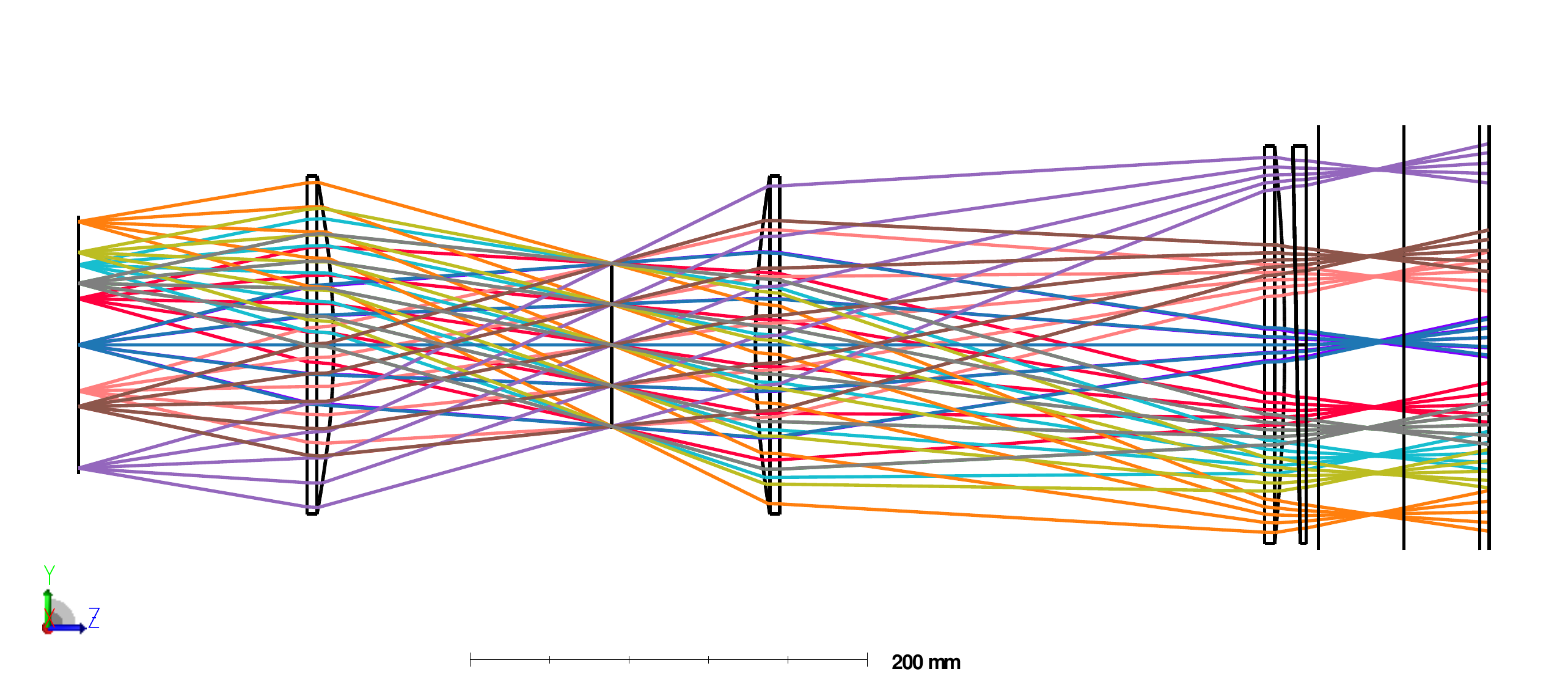}
	\caption{Camera optical layout for the TMA LAT showing: the three silicon lenses, the Lyot stop and the alumina prism. From left to right: Focal plane, L3, Lyot stop, L2, L1, Alumina wedge and filter stack. Rays are traced in the time reverse sense, ie: from left to right. Colors indicate different detector positions in the detector focal plane. The focal plane has $130 \, \rm mm$ in diameter, L1 measures $200 \, \rm mm$, while L2 and L3 measure  $170\, \rm mm$ in diameter. The crossed-Dragone camera results in a very similar layout.}
	\label{fig:camera_layout}
\end{figure}

\begin{algorithm}[htb]
\begin{enumerate}
\caption{Merit function pseudo-code to be used in the optimization of the TMA and CD camera designs. Steps in this list are performed sequentially and updated in each evaluation during the optimization. Each line represents one term in the sum shown in Equation \ref{mf:eq}. Each term is  composed of multiple ray evaluations in Zemax. In practice, the merit function to be optimized is composed of hundreds of terms.}
\label{alg:mf}
\centering
\item Illuminate Lyot stop evenly
\item Limits on lens thickness and camera total length
\item Constrain wedge slope sign
\item Constrain ellipticity on a screen past L3
\item All rays within L1
\item All rays within L2
\item All rays within L3
\item $f/\#$ constraint
\item $f/\#$ ellipticity constraint
\item Chief ray focal plane angle constraint
\item Prism wedge tilt and clocking
\item Minimize RMS spot size

\end{enumerate}
\end{algorithm}

The five-meter TMA can be manufactured with gap-less mirrors. These mirrors can be machined in halves and bolted together to form reflectors which present a continuous and smooth surface to incoming light. This smooth surface avoids diffraction sidelobes that originate from the gaps in the paneled surface, which has been historically used in the manufacture of large-aperture telescopes for CMB studies (and are present in the CHLAT design). The telescope structure is being designed to meet the stability requirements determined from optical tolerancing analysis, as well as from the expected environmental and operational loads. Optomechanical performance will be discussed in a future publication.

\section{Camera Arrays}
\label{sec:implementation}

\begin{table}[tb]
\centering
\begin{tabular}{c|cccc}
	component & $R^{-1}\,\rm [mm^{-1}]$ & c  & $t_{j,{j+1}} \, \rm [mm]$\\
	\hline
	L1 & $-9.68\times 10 ^{-4}$ & -1  & 271\\
	L2 & $1.99\times 10^{-3}$ & -1  & 78.5\\
	stop & -- & -- & 152\\
	L3 & $-2.29\times 10^{-3}$ & -1  & 120\\
\end{tabular}

\caption{Initial values to be used with the merit function described by Algorithm \ref{alg:mf} to optimize for a suitable optical solution for the camera lenses. Rows indicate an optical component, $R$ is the radius of curvature, $k$ is the conic constant and $t_{j, j+1}$ is the physical distance that separates the current surface from the next. The coordinate system is defined in the time reverse sense as shown in Figure \ref{fig:camera_layout}, such that the radii of curvature for L1 and L3 are negative, while L2 curves in the opposite direction. In the TMA system, a camera solution is defined by 10 variables, while in the biconic CD camera, the optimization solves for 16 variables. The wedge tilt and clocking angle are also left as variables but they are optimized again after the lens solution has been reached with a more detailed merit function.}
\label{tab:initialvals}
\end{table}

\begin{figure}[p]
    \centering
    \begin{subfigure}{.4\textwidth}
        \centering
	    \includegraphics[width=.7\linewidth]{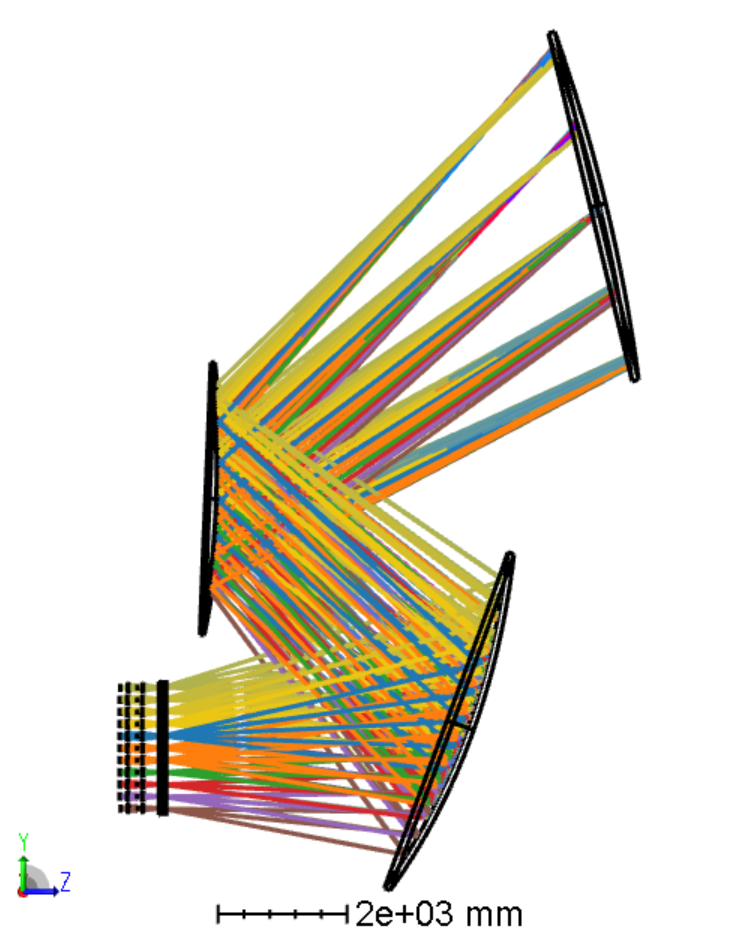}
    \end{subfigure}
  \begin{subfigure}{.5\textwidth}
      \centering
      \includegraphics[width=\linewidth]{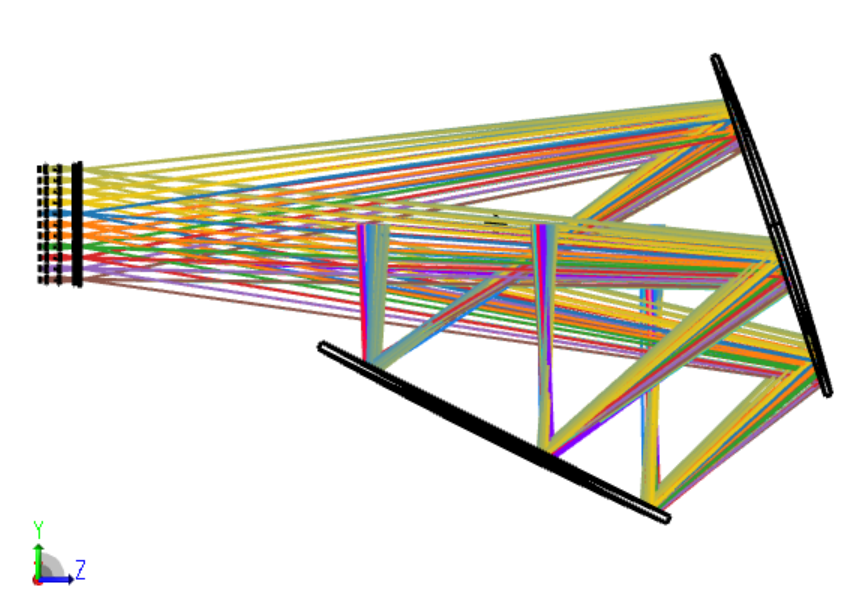}
\end{subfigure}%

    \centering
    \begin{subfigure}{.48\textwidth}
        \centering
        \includegraphics[width=.8\linewidth]{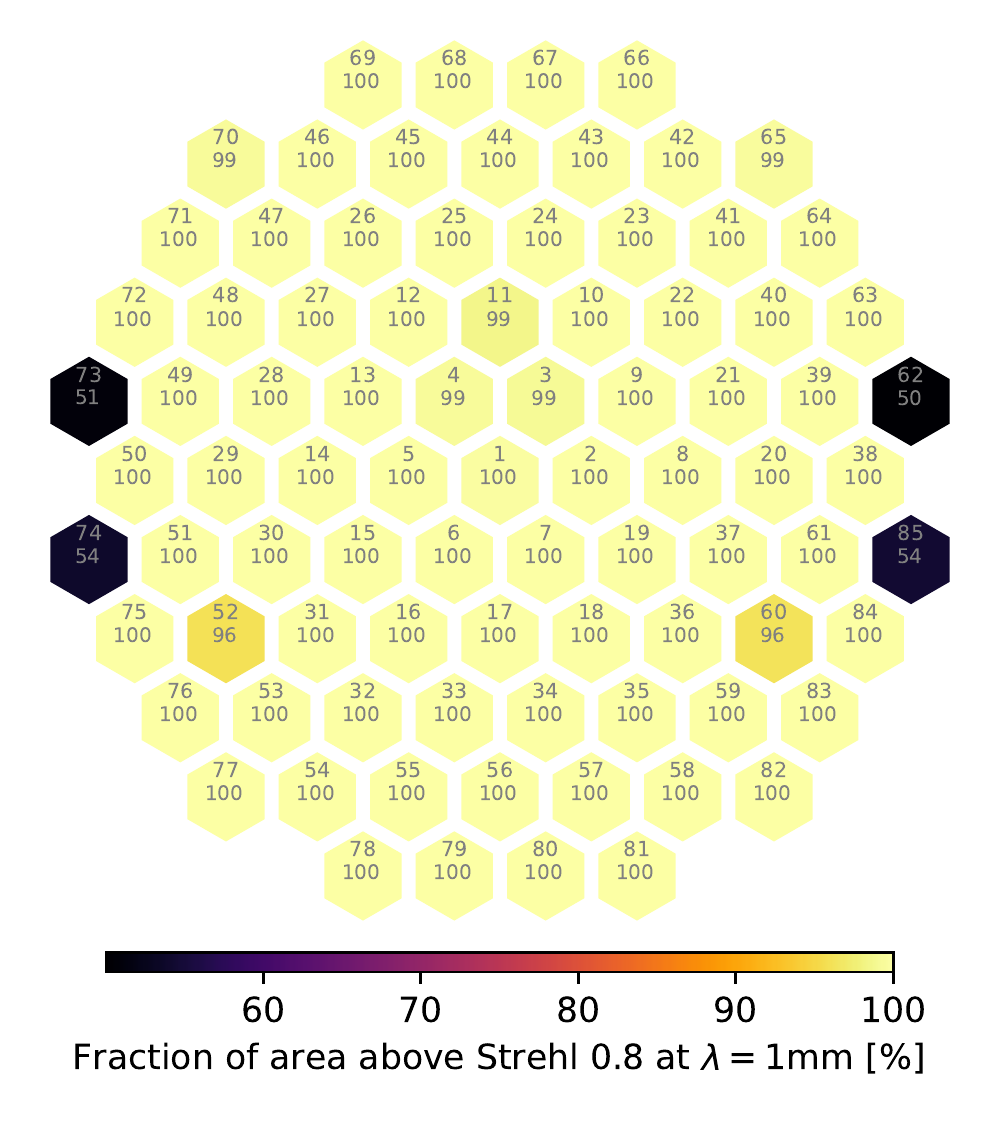}
    \end{subfigure}
  \begin{subfigure}{.48\textwidth}
      \centering
      \includegraphics[width=.8\linewidth]{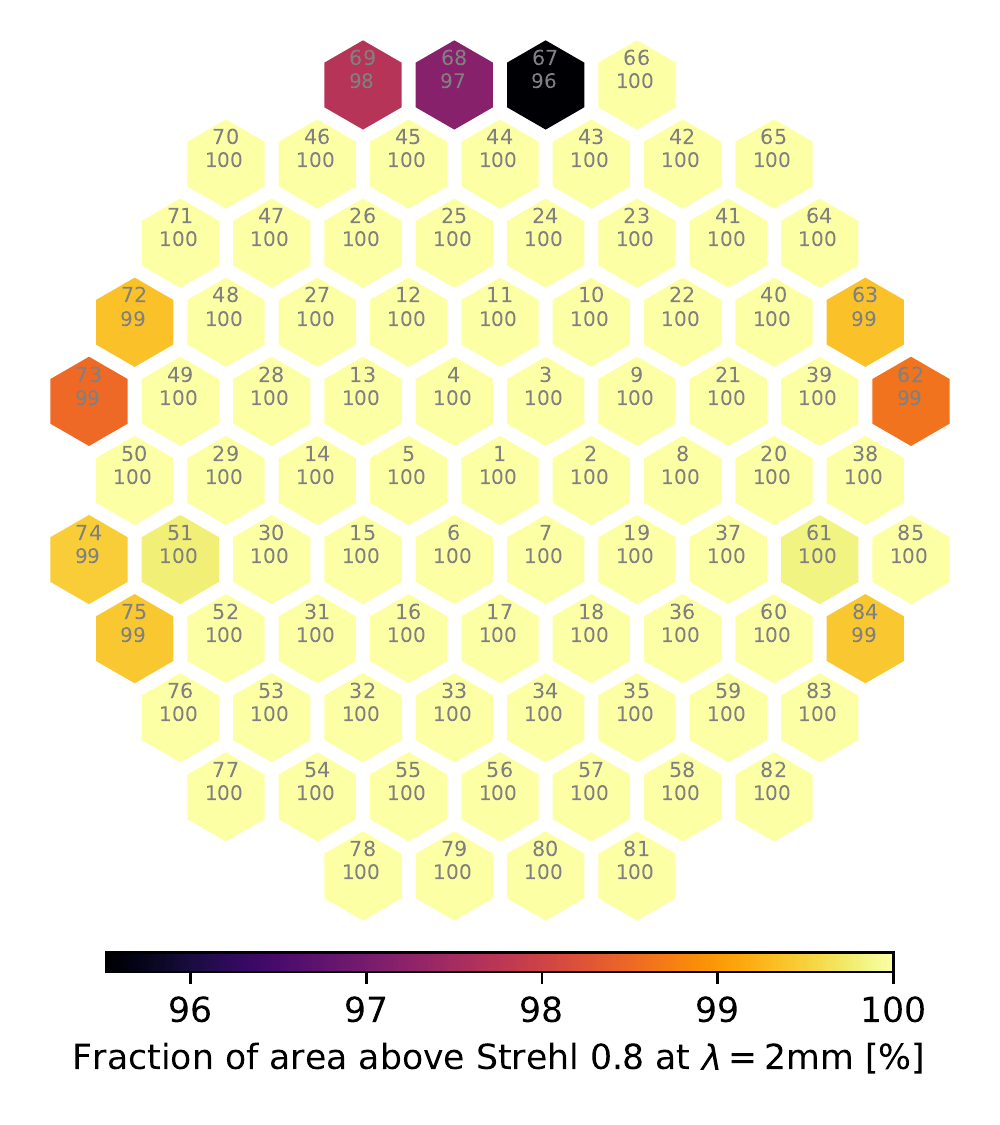}
\end{subfigure}%
    \caption{Top: Ray trace for the two telescope concepts with their arrays of cameras. Upper Left: TMA with 85 cameras using only two types of lens prescriptions. Upper Right: crossed-Dragone design with 85 cameras optimized (biconic L2) individually.
Bottom: Diffraction-limited (Strehl ratio $> 0.8$) focal plane fraction (percent units) for all 85 cameras. The footprint of this hexagonal array covers the same area in both telescopes. 
Bottom Left: Strehl ratio coverage for the TMA array of 85 cameras using two groups of optical prescriptions and individual alumina wedges. Plot shows diffraction-limited performance (Strehl ratio $>$ 0.8) at $1\, \rm mm$ in 81 of the 85 cameras. Diffraction-limited performance for all cameras is achieved at $2\, \rm mm$ of wavelength. Bottom Right: Image quality coverage for the crossed-Dragone optical design with individual lens prescriptions for the 85 cameras. Map shows the existence of solutions at $2\, \rm mm$ with diffraction-limited performance using a biconic lens L2 to correct astigmatism. This solution set is being evaluated in order to select a minimal set of groups that show acceptable image quality while minimizing fabrication complexity.}
    \label{fig:mosaicImgQualDiagrams}
\end{figure}

The concept for the CMB-S4 array of cameras consists of 85 cameras arranged in a hexagonal pattern covering different bands. Each camera is composed of a tube that holds silicon lenses (``optics tubes'' for short), and filters. These optics tubes are housed in a single cryostat that cools the system. The optical prescription for these optics tubes can be different among cameras, depending on the optical properties of the LAT. Because the LATs were designed to have a similar f-number and focal plane size, the footprint of the camera array is arranged to be identical in both telescope concepts (SPLAT and CHLAT); therefore they can use the same camera to camera separation and diameter. This simplifies the design greatly and it enables shared features among mechanical designs. Figure \ref{fig:hexes} shows the distribution of cameras and bands in this hexagonal pattern. Figure \ref{fig:camera_layout} shows the layout of one camera system and Figure \ref{fig:mosaicImgQualDiagrams} (top) shows the 85 camera system layout as intended to work with the LAT optics.

Each camera is composed of three plano-convex lenses. These lenses are labeled  L1, L2 and L3 in the direction of propagation of light. The TMA has minimal astigmatism, which allows the three lenses to be radially symmetric and their curved surface is described by \begin{equation}
z(r) = \frac{(r/R)^2}{1 + \sqrt{1-(1+k)(r/R)^2}},
\end{equation} where $R$ is the radius of curvature, $k$ is the conic constant ($k<-1$ for hyperbolas, $k=0$ for parabolas, $0<k<-1$ for ellipses, $k=0$ for spheres and $k>0$ for oblate ellipsoids) and $r$ is the radial distance from the axis of symmetry of the lens. Astigmatism in the crossed-Dragone telescope can be corrected with the use of one biconic surface in the lens that is closest to the Lyot stop (L2). A biconic lens has two radii of curvature in two orthogonal directions and has a shape given by\begin{equation}
z(x, y) = \frac{(x/R_x)^2 + (y/R_y)^2}{1 + \sqrt{1-(1+k_x)(x/R_x)^2 - (1+k_y)(y/R_y)^2 }},
\end{equation} where $x$ and $y$ are two coordinates perpendicular to the axis of the lens, the sub-index denotes parameters in the $x$ or $y$ direction. The three lenses in the camera are made of silicon ($n=3.38$), and before L1 there is a alumina ($n=3.14$) wedge that corrects for the curvature of the telescope focal surface. The tilt in the surface of this prism and the clocking angle are optimized jointly while finding the shapes of the lenses.

The design of the three-lens camera system is best performed in the time-reverse sense (focal plane $\rightarrow$ L3$\rightarrow$ Lyot $\rightarrow$ L2 $\rightarrow$ L1 $\rightarrow$ Telescope), as it is numerically easier to set the pupil-defining aperture close to the start of the ray trace than at the end. This time-reverse design requires time-reversing the large-aperture telescope\cite{zmxmanual}. We start the optimization with lens parameter values that roughly focus light into a 6 inch focal plane (see Table \ref{tab:initialvals} for initial values), while re-imaging the primary in an intermediate surface (Lyot stop) between L2 and L3. 

We define a merit function as the weighted quadratic sum of the differences between the current design and target values given by \begin{equation}
MF = \frac{\sum W_j (V_j - T_j)^2}{\sum W_j},
\label{mf:eq}
\end{equation}
 where $W_j$ are weights that are used to set the importance of each term, $V_j$ are the values obtained by evaluating the current design and $T_j$ a target value. The list of values that are evaluated sequentially  is shown schematically as pseudo-code in Algorithm \ref{alg:mf} (the actual merit function has hundreds of terms). The optimization consists of minimizing the quantity $MF$ using numerical techniques to obtain the best lens shapes and distances considering these constraints. As shown in Algorithm \ref{alg:mf}, we use numerical constraints to keep all rays within the limits of their apertures (diameters of $200\, \rm mm$ for L1 and $170\, \rm mm$ for L2 and L3); we also use constraints that limit the telecentricity (angle between the normal and the chief ray) at the focal plane to be lower than $2.5\, \rm deg$, and a constraint for the maximum f-number asymmetry of 12\% evaluated in an imaginary screen beyond L3. Finally we use the R.M.S. spot size  (see Zemax manual\cite{zmxmanual}) to optimize for optical quality. The optimization was performed in Zemax Optics Studio using the damped least squares method, which is accessed via the Python API. We optimized sequentially all 85 cameras to demonstrate that suitable solutions exist for each camera. After we found suitable solutions for all the cameras, we used these individual solutions to evaluate the system with identical lens prescriptions. We make exhaustive evaluations of camera performance using these 85 solutions copied to all cameras to assess how many cameras can have the same optical prescription.

Using the methods described in this section we found that a satisfactory solution (which gives diffraction-limited performance in all but four cameras at $1\, \rm mm$) can be found for the TMA if we define two groups of cameras (details of this solution will be presented in Gallardo 2022 in prep.\cite{2022Gallardo}). For the crossed-Dragone design, we have shown that astigmatism can be cancelled out at 2 mm by the use of a biconic lens L2 optimizing all cameras individually, which demonstrates the existence of solutions. We are currently searching for grouped solutions for this design. Figure \ref{fig:mosaicImgQualDiagrams} (bottom) shows the fraction of the focal plane covered with diffraction-limited performance (Strehl $>$ 0.8) for the TMA (two groups of optical prescriptions) and CD (individual camera prescriptions) design at $1\, \rm mm$ and $2\, \rm mm$ in wavelength, respectively. The iterative process of replicating camera prescriptions and finding suitable groups is being carried out for the crossed-Dragone telescope design cameras to find a grouped solution that minimizes fabrication complexity with a biconic prescription. Opto-mechanical checks are being carried out to define manufacturability constraints that will inform the optical prescription as the design converges. Table \ref{tab:currentstatus} shows a summary of the current status of the optical design.

\begin{table}
    \centering
    \begin{tabular}{c|c|c|c}
        Concept & Individual Camera Solution Found  & Grouped Lenses Solution Found & Opto-Mechanical\\
        \hline
        SPLAT & Yes & Yes & In progress\\
        CHLAT & Yes & In progress & In progress
    \end{tabular}
    \caption{Current progress in the design of the 85-camera system. Second column (Individual Camera Solution Found) describes the status of the search of a solution for all three lenses and Lyot stop that is unique for each one of the 85 cameras. This step in the design is used to prove the existence of solutions and to gather information about the parameter space. Third column (Grouped Lenses Solution Found)  shows if the solution found in the first step has been replaced with fewer groups of identical lenses. This step is done to minimize fabrication complexity. The fourth column (Opto-mechanical) shows progress on the integration between the optical and mechanical design.}
    \label{tab:currentstatus}
\end{table}

\section{Conclusion}
\label{sec:conclusion}

We presented the optical design concept and current status for the conceptual design of CMB-S4. This optical concept is composed of two types of large-aperture telescopes, a CD and a TMA, with their respective arrays of 85 cameras. We have identified two groups of optical prescriptions that fully populate the 85 cameras in the TMA design, and we are working on identifying groups of optical solutions for the biconic CD cameras. A prototype primary mirror of the TMA is complete, demonstrating the joining technology. We have discussed the computational methods used to generate families of solutions given mechanical constraints for the 85 camera arrays. Opto-mechanical iterations are being carried out to converge on an optical solution of this design using the tools that we have presented here. Diffraction properties in the point spread function will be evaluated when the optical design prescription converges. The optical concept we have presented will allow to populate the focal planes of these large-aperture telescopes with 85 cameras to achieve an unprecedented increase in mapping speed in large-aperture CMB experiments.

\acknowledgments

The authors acknowledge with heavy hearts the invaluable contribution of Richard Hills (1945-2022), who passed away during the preparation of this document. Richard optimized the TMA optics design and made a significant contribution to the design of the camera array presented here. He patiently tutored PAG in the art and science of optical design. We will sorely miss him.

CMB-S4 is supported by the Director, Office of Science, Office of High Energy Physics of the U.S. Department of Energy under Contract No.DE–AC02–05CH11231; by the National Energy Research Scientific Computing Center, a DOE Office of Science User Facility under the same contract; and by the Divisions of Physics and Astronomical Sciences and the Office of Polar Programs of the U.S. National Science Foundation under Mid-Scale Research Infrastructure award OPP-1935892. Considerable additional support is provided by the many CMB-S4 team members and their institutions.

Work supported by the Fermi National Accelerator Laboratory, managed and operated by Fermi Research Alliance, LLC under Contract No. DE-AC02-07CH11359 with the U.S. Department of Energy. The U.S. Government retains and the publisher, by accepting the article for publication, acknowledges that the U.S. Government retains a non-exclusive, paid-up, irrevocable, world-wide license to publish or reproduce the published form of this manuscript, or allow others to do so, for U.S. Government purposes.

\bibliography{biblio} 

\begin{thebibliography}{10}

\bibitem{sciencecase}
Abazajian, K., Addison, G., Adshead, P., Ahmed, Z., Allen, S.~W., Alonso, D.,
  Alvarez, M., Anderson, A., Arnold, K.~S., Baccigalupi, C., Bailey, K.,
  Barkats, D., Barron, D., Barry, P.~S., Bartlett, J.~G., Thakur, R.~B.,
  Battaglia, N., Baxter, E., Bean, R., Bebek, C., Bender, A.~N., Benson, B.~A.,
  Berger, E., Bhimani, S., Bischoff, C.~A., Bleem, L., Bocquet, S., Boddy, K.,
  Bonato, M., Bond, J.~R., Borrill, J., Bouchet, F.~R., Brown, M.~L., Bryan,
  S., Burkhart, B., Buza, V., Byrum, K., Calabrese, E., Calafut, V., Caldwell,
  R., Carlstrom, J.~E., Carron, J., Cecil, T., Challinor, A., Chang, C.~L.,
  Chinone, Y., Cho, H.-M.~S., Cooray, A., Crawford, T.~M., Crites, A.,
  Cukierman, A., Cyr-Racine, F.-Y., de~Haan, T., de~Zotti, G., Delabrouille,
  J., Demarteau, M., Devlin, M., Di~Valentino, E., Dobbs, M., Duff, S.,
  Duivenvoorden, A., Dvorkin, C., Edwards, W., Eimer, J., Errard, J.,
  Essinger-Hileman, T., Fabbian, G., Feng, C., Ferraro, S., Filippini, J.~P.,
  Flauger, R., Flaugher, B., Fraisse, A.~A., Frolov, A., Galitzki, N., Galli,
  S., Ganga, K., Gerbino, M., Gilchriese, M., Gluscevic, V., Green, D., Grin,
  D., Grohs, E., Gualtieri, R., Guarino, V., Gudmundsson, J.~E., Habib, S.,
  Haller, G., Halpern, M., Halverson, N.~W., Hanany, S., Harrington, K.,
  Hasegawa, M., Hasselfield, M., Hazumi, M., Heitmann, K., Henderson, S.,
  Henning, J.~W., Hill, J.~C., Hlozek, R., Holder, G., Holzapfel, W., Hubmayr,
  J., Huffenberger, K.~M., Huffer, M., Hui, H., Irwin, K., Johnson, B.~R.,
  Johnstone, D., Jones, W.~C., Karkare, K., Katayama, N., Kerby, J., Kernovsky,
  S., Keskitalo, R., Kisner, T., Knox, L., Kosowsky, A., Kovac, J., Kovetz,
  E.~D., Kuhlmann, S., Kuo, C.-l., Kurita, N., Kusaka, A., Lahteenmaki, A.,
  Lawrence, C.~R., Lee, A.~T., Lewis, A., Li, D., Linder, E., Loverde, M.,
  Lowitz, A., Madhavacheril, M.~S., Mantz, A., Matsuda, F., Mauskopf, P.,
  McMahon, J., McQuinn, M., Meerburg, P.~D., Melin, J.-B., Meyers, J., Millea,
  M., Mohr, J., Moncelsi, L., Mroczkowski, T., Mukherjee, S., Münchmeyer, M.,
  Nagai, D., Nagy, J., Namikawa, T., Nati, F., Natoli, T., Negrello, M.,
  Newburgh, L., Niemack, M.~D., Nishino, H., Nordby, M., Novosad, V., O'Connor,
  P., Obied, G., Padin, S., Pandey, S., Partridge, B., Pierpaoli, E., Pogosian,
  L., Pryke, C., Puglisi, G., Racine, B., Raghunathan, S., Rahlin, A.,
  Rajagopalan, S., Raveri, M., Reichanadter, M., Reichardt, C.~L., Remazeilles,
  M., Rocha, G., Roe, N.~A., Roy, A., Ruhl, J., Salatino, M., Saliwanchik, B.,
  Schaan, E., Schillaci, A., Schmittfull, M.~M., Scott, D., Sehgal, N.,
  Shandera, S., Sheehy, C., Sherwin, B.~D., Shirokoff, E., Simon, S.~M.,
  Slosar, A., Somerville, R., Spergel, D., Staggs, S.~T., Stark, A., Stompor,
  R., Story, K.~T., Stoughton, C., Suzuki, A., Tajima, O., Teply, G.~P.,
  Thompson, K., Timbie, P., Tomasi, M., Treu, J.~I., Tristram, M., Tucker, G.,
  Umiltà, C., van Engelen, A., Vieira, J.~D., Vieregg, A.~G., Vogelsberger,
  M., Wang, G., Watson, S., White, M., Whitehorn, N., Wollack, E.~J., Wu, W.
  L.~K., Xu, Z., Yasini, S., Yeck, J., Yoon, K.~W., Young, E., and Zonca, A.,
  ``Cmb-s4 science case, reference design, and project plan,'' (2019).

\bibitem{sciencebook}
Abazajian, K.~N., Adshead, P., Ahmed, Z., Allen, S.~W., Alonso, D., Arnold,
  K.~S., Baccigalupi, C., Bartlett, J.~G., Battaglia, N., Benson, B.~A.,
  Bischoff, C.~A., Borrill, J., Buza, V., Calabrese, E., Caldwell, R.,
  Carlstrom, J.~E., Chang, C.~L., Crawford, T.~M., Cyr-Racine, F.-Y.,
  De~Bernardis, F., de~Haan, T., Alighieri, S. d.~S., Dunkley, J., Dvorkin, C.,
  Errard, J., Fabbian, G., Feeney, S., Ferraro, S., Filippini, J.~P., Flauger,
  R., Fuller, G.~M., Gluscevic, V., Green, D., Grin, D., Grohs, E., Henning,
  J.~W., Hill, J.~C., Hlozek, R., Holder, G., Holzapfel, W., Hu, W.,
  Huffenberger, K.~M., Keskitalo, R., Knox, L., Kosowsky, A., Kovac, J.,
  Kovetz, E.~D., Kuo, C.-L., Kusaka, A., Jeune, M.~L., Lee, A.~T., Lilley, M.,
  Loverde, M., Madhavacheril, M.~S., Mantz, A., Marsh, D. J.~E., McMahon, J.,
  Meerburg, P.~D., Meyers, J., Miller, A.~D., Munoz, J.~B., Nguyen, H.~N.,
  Niemack, M.~D., Peloso, M., Peloton, J., Pogosian, L., Pryke, C., Raveri, M.,
  Reichardt, C.~L., Rocha, G., Rotti, A., Schaan, E., Schmittfull, M.~M.,
  Scott, D., Sehgal, N., Shandera, S., Sherwin, B.~D., Smith, T.~L., Sorbo, L.,
  Starkman, G.~D., Story, K.~T., van Engelen, A., Vieira, J.~D., Watson, S.,
  Whitehorn, N., and Wu, W. L.~K., ``Cmb-s4 science book, first edition,''
  (2016).

\bibitem{technologybook}
Abitbol, M.~H., Ahmed, Z., Barron, D., Thakur, R.~B., Bender, A.~N., Benson,
  B.~A., Bischoff, C.~A., Bryan, S.~A., Carlstrom, J.~E., Chang, C.~L., Chuss,
  D.~T., Crowley, K.~T., Cukierman, A., de~Haan, T., Dobbs, M.,
  Essinger-Hileman, T., Filippini, J.~P., Ganga, K., Gudmundsson, J.~E.,
  Halverson, N.~W., Hanany, S., Henderson, S.~W., Hill, C.~A., Ho, S.-P.~P.,
  Hubmayr, J., Irwin, K., Jeong, O., Johnson, B.~R., Kernasovskiy, S.~A.,
  Kovac, J.~M., Kusaka, A., Lee, A.~T., Maria, S., Mauskopf, P., McMahon,
  J.~J., Moncelsi, L., Nadolski, A.~W., Nagy, J.~M., Niemack, M.~D., O'Brient,
  R.~C., Padin, S., Parshley, S.~C., Pryke, C., Roe, N.~A., Rostem, K., Ruhl,
  J., Simon, S.~M., Staggs, S.~T., Suzuki, A., Switzer, E.~R., Tajima, O.,
  Thompson, K.~L., Timbie, P., Tucker, G.~S., Vieira, J.~D., Vieregg, A.~G.,
  Westbrook, B., Wollack, E.~J., Yoon, K.~W., Young, K.~S., and Young, E.~Y.,
  ``Cmb-s4 technology book, first edition,'' (2017).

\bibitem{whitepaper}
Abazajian, K., Abdulghafour, A., Addison, G.~E., Adshead, P., Ahmed, Z.,
  Ajello, M., Akerib, D., Allen, S.~W., Alonso, D., Alvarez, M., Amin, M.~A.,
  Amiri, M., Anderson, A., Ansarinejad, B., Archipley, M., Arnold, K.~S.,
  Ashby, M., Aung, H., Baccigalupi, C., Baker, C., Bakshi, A., Bard, D.,
  Barkats, D., Barron, D., Barry, P.~S., Bartlett, J.~G., Barton, P., Thakur,
  R.~B., Battaglia, N., Beall, J., Bean, R., Beck, D., Belkner, S., Benabed,
  K., Bender, A.~N., Benson, B.~A., Besuner, B., Bethermin, M., Bhimani, S.,
  Bianchini, F., Biquard, S., Birdwell, I., Bischoff, C.~A., Bleem, L., Bocaz,
  P., Bock, J.~J., Bocquet, S., Boddy, K.~K., Bond, J.~R., Borrill, J.,
  Bouchet, F.~R., Brinckmann, T., Brown, M.~L., Bryan, S., Buza, V., Byrum, K.,
  Calabrese, E., Calafut, V., Caldwell, R., Carlstrom, J.~E., Carron, J.,
  Cecil, T., Challinor, A., Chan, V., Chang, C.~L., Chapman, S., Charles, E.,
  Chauvin, E., Cheng, C., Chesmore, G., Cheung, K., Chinone, Y., Chluba, J.,
  Cho, H.-M.~S., Choi, S., Clancy, J., Clark, S., Cooray, A., Coppi, G.,
  Corlett, J., Coulton, W., Crawford, T.~M., Crites, A., Cukierman, A.,
  Cyr-Racine, F.-Y., Dai, W.-M., Daley, C., Dart, E., Daues, G., de~Haan, T.,
  Deaconu, C., Delabrouille, J., Derylo, G., Devlin, M., Di~Valentino, E.,
  Dierickx, M., Dober, B., Doriese, R., Duff, S., Dutcher, D., Dvorkin, C.,
  Dünner, R., Eftekhari, T., Eimer, J., Bouhargani, H.~E., Elleflot, T.,
  Emerson, N., Errard, J., Essinger-Hileman, T., Fabbian, G., Fanfani, V.,
  Fasano, A., Feng, C., Ferraro, S., Filippini, J.~P., Flauger, R., Flaugher,
  B., Fraisse, A.~A., Frisch, J., Frolov, A., Galitzki, N., Gallardo, P.~A.,
  Galli, S., Ganga, K., Gerbino, M., Giannakopoulos, C., Gilchriese, M.,
  Gluscevic, V., Goeckner-Wald, N., Goldfinger, D., Green, D., Grimes, P.,
  Grin, D., Grohs, E., Gualtieri, R., Guarino, V., Gudmundsson, J.~E., Gullett,
  I., Guns, S., Habib, S., Haller, G., Halpern, M., Halverson, N.~W., Hanany,
  S., Hand, E., Harrington, K., Hasegawa, M., Hasselfield, M., Hazumi, M.,
  Heitmann, K., Henderson, S., Hensley, B., Herbst, R., Hervias-Caimapo, C.,
  Hill, J.~C., Hills, R., Hivon, E., Hlozek, R., Ho, A., Holder, G., Hollister,
  M., Holzapfel, W., Hood, J., Hotinli, S., Hryciuk, A., Hubmayr, J.,
  Huffenberger, K.~M., Hui, H., nez, R.~I., Ibitoye, A., Ikape, M., Irwin, K.,
  Jacobus, C., Jeong, O., Johnson, B.~R., Johnstone, D., Jones, W.~C., Joseph,
  J., Jost, B., Kang, J.~H., Kaplan, A., Karkare, K.~S., Katayama, N.,
  Keskitalo, R., King, C., Kisner, T., Klein, M., Knox, L., Koopman, B.~J.,
  Kosowsky, A., Kovac, J., Kovetz, E.~D., Krolewski, A., Kubik, D., Kuhlmann,
  S., Kuo, C.-L., Kusaka, A., Lähteenmäki, A., Lau, K., Lawrence, C.~R., Lee,
  A.~T., Legrand, L., Leitner, M., Leloup, C., Lewis, A., Li, D., Linder, E.,
  Liodakis, I., Liu, J., Long, K., Louis, T., Loverde, M., Lowry, L., Lu, C.,
  Lubin, P., Ma, Y.-Z., Maccarone, T., Madhavacheril, M.~S., Maldonado, F.,
  Mantz, A., Marques, G., Matsuda, F., Mauskopf, P., May, J., McCarrick, H.,
  McCracken, K., McMahon, J., Meerburg, P.~D., Melin, J.-B., Menanteau, F.,
  Meyers, J., Millea, M., Miranda, V., Mitchell, D., Mohr, J., Moncelsi, L.,
  Monzani, M.~E., Moshed, M., Mroczkowski, T., Mukherjee, S., Münchmeyer, M.,
  Nagai, D., Nagarajappa, C., Nagy, J., Namikawa, T., Nati, F., Natoli, T.,
  Nerval, S., Newburgh, L., Nguyen, H., Nichols, E., Nicola, A., Niemack,
  M.~D., Nord, B., Norton, T., Novosad, V., O'Brient, R., Omori, Y., Orlando,
  G., Osherson, B., Osten, R., Padin, S., Paine, S., Partridge, B., Patil, S.,
  Petravick, D., Petroff, M., Pierpaoli, E., Pilleux, M., Pogosian, L., Prabhu,
  K., Pryke, C., Puglisi, G., Racine, B., Raghunathan, S., Rahlin, A., Raveri,
  M., Reese, B., Reichardt, C.~L., Remazeilles, M., Rizzieri, A., Rocha, G.,
  Roe, N.~A., Rotermund, K., Roy, A., Ruhl, J.~E., Saba, J., Sailer, N.,
  Salatino, M., Saliwanchik, B., Sapozhnikov, L., Rao, M.~S., Saunders, L.,
  Schaan, E., Schillaci, A., Schmitt, B., Scott, D., Sehgal, N., Shandera, S.,
  Sherwin, B.~D., Shirokoff, E., Shiu, C., Simon, S.~M., Singari, B., Slosar,
  A., Spergel, D., Germaine, T.~S., Staggs, S.~T., Stark, A.~A., Starkman,
  G.~D., Steinbach, B., Stompor, R., Stoughton, C., Suzuki, A., Tajima, O.,
  Tandoi, C., Teply, G.~P., Thayer, G., Thompson, K., Thorne, B., Timbie, P.,
  Tomasi, M., Trendafilova, C., Tristram, M., Tucker, C., Tucker, G., Umiltà,
  C., van Engelen, A., van Marrewijk, J., Vavagiakis, E.~M., Vergès, C.,
  Vieira, J.~D., Vieregg, A.~G., Wagoner, K., Wallisch, B., Wang, G., Wang,
  G.-J., Watson, S., Watts, D., Weaver, C., Wenzl, L., Westbrook, B., White,
  M., Whitehorn, N., Wiedlea, A., Williams, P., Wilson, R., Winch, H., Wollack,
  E.~J., Wu, W. L.~K., Xu, Z., Yefremenko, V.~G., Yu, C., Zegeye, D., Zivick,
  J., and Zonca, A., ``Snowmass 2021 cmb-s4 white paper,'' (2022).

\bibitem{2021ApOpt..60..823G}
{Gudmundsson}, J.~E., {Gallardo}, P.~A., {Puddu}, R., {Dicker}, S.~R., {Adler},
  A.~E., {Ali}, A.~M., {Bazarko}, A., {Chesmore}, G.~E., {Coppi}, G.,
  {Cothard}, N.~F., {Dachlythra}, N., {Devlin}, M., {D{\"u}nner}, R.,
  {Fabbian}, G., {Galitzki}, N., {Golec}, J.~E., {Patty Ho}, S.-P., {Hargrave},
  P.~C., {Kofman}, A.~M., {Lee}, A.~T., {Limon}, M., {Matsuda}, F.~T.,
  {Mauskopf}, P.~D., {Moodley}, K., {Nati}, F., {Niemack}, M.~D.,
  {Orlowski-Scherer}, J., {Page}, L.~A., {Partridge}, B., {Puglisi}, G.,
  {Reichardt}, C.~L., {Sierra}, C.~E., {Simon}, S.~M., {Teply}, G.~P.,
  {Tucker}, C., {Wollack}, E.~J., {Xu}, Z., and {Zhu}, N., ``{The Simons
  Observatory: modeling optical systematics in the Large Aperture Telescope},''
  {\em Applied Optics}~{\bf 60},  823 (Feb. 2021).

\bibitem{ApOpt..57.2314P}
{Padin}, S., ``{Three-mirror anastigmat for cosmic microwave background
  observations},'' {\em Applied Optics}~{\bf 57},  2314 (Mar. 2018).

\bibitem{2018SPIE10700E..41P}
{Parshley}, S.~C., {Niemack}, M., {Hills}, R., {Dicker}, S.~R., {D{\"u}nner},
  R., {Erler}, J., {Gallardo}, P.~A., {Gudmundsson}, J.~E., {Herter}, T.,
  {Koopman}, B.~J., {Limon}, M., {Matsuda}, F.~T., {Mauskopf}, P., {Riechers},
  D.~A., {Stacey}, G.~J., and {Vavagiakis}, E.~M., ``{The optical design of the
  six-meter CCAT-prime and Simons Observatory telescopes},'' in [{\em
  Ground-based and Airborne Telescopes VII}{\nolinebreak\hspace{0.1em}]},
  {Marshall}, H.~K. and {Spyromilio}, J., eds., {\em Society of Photo-Optical
  Instrumentation Engineers (SPIE) Conference Series} {\bf 10700},  1070041
  (July 2018).

\bibitem{2016JLTP..184..772H}
{Henderson}, S.~W., {Allison}, R., {Austermann}, J., {Baildon}, T.,
  {Battaglia}, N., {Beall}, J.~A., {Becker}, D., {De Bernardis}, F., {Bond},
  J.~R., {Calabrese}, E., {Choi}, S.~K., {Coughlin}, K.~P., {Crowley}, K.~T.,
  {Datta}, R., {Devlin}, M.~J., {Duff}, S.~M., {Dunkley}, J., {D{\"u}nner}, R.,
  {van Engelen}, A., {Gallardo}, P.~A., {Grace}, E., {Hasselfield}, M.,
  {Hills}, F., {Hilton}, G.~C., {Hincks}, A.~D., {Hlozek}, R., {Ho}, S.~P.,
  {Hubmayr}, J., {Huffenberger}, K., {Hughes}, J.~P., {Irwin}, K.~D.,
  {Koopman}, B.~J., {Kosowsky}, A.~B., {Li}, D., {McMahon}, J., {Munson}, C.,
  {Nati}, F., {Newburgh}, L., {Niemack}, M.~D., {Niraula}, P., {Page}, L.~A.,
  {Pappas}, C.~G., {Salatino}, M., {Schillaci}, A., {Schmitt}, B.~L., {Sehgal},
  N., {Sherwin}, B.~D., {Sievers}, J.~L., {Simon}, S.~M., {Spergel}, D.~N.,
  {Staggs}, S.~T., {Stevens}, J.~R., {Thornton}, R., {Van Lanen}, J.,
  {Vavagiakis}, E.~M., {Ward}, J.~T., and {Wollack}, E.~J., ``{Advanced ACTPol
  Cryogenic Detector Arrays and Readout},'' {\em Journal of Low Temperature
  Physics}~{\bf 184},  772--779 (Aug. 2016).

\bibitem{2014SPIE.9153E..1PB}
{Benson}, B.~A., {Ade}, P.~A.~R., {Ahmed}, Z., {Allen}, S.~W., {Arnold}, K.,
  {Austermann}, J.~E., {Bender}, A.~N., {Bleem}, L.~E., {Carlstrom}, J.~E.,
  {Chang}, C.~L., {Cho}, H.~M., {Cliche}, J.~F., {Crawford}, T.~M.,
  {Cukierman}, A., {de Haan}, T., {Dobbs}, M.~A., {Dutcher}, D., {Everett}, W.,
  {Gilbert}, A., {Halverson}, N.~W., {Hanson}, D., {Harrington}, N.~L.,
  {Hattori}, K., {Henning}, J.~W., {Hilton}, G.~C., {Holder}, G.~P.,
  {Holzapfel}, W.~L., {Irwin}, K.~D., {Keisler}, R., {Knox}, L., {Kubik}, D.,
  {Kuo}, C.~L., {Lee}, A.~T., {Leitch}, E.~M., {Li}, D., {McDonald}, M.,
  {Meyer}, S.~S., {Montgomery}, J., {Myers}, M., {Natoli}, T., {Nguyen}, H.,
  {Novosad}, V., {Padin}, S., {Pan}, Z., {Pearson}, J., {Reichardt}, C.,
  {Ruhl}, J.~E., {Saliwanchik}, B.~R., {Simard}, G., {Smecher}, G., {Sayre},
  J.~T., {Shirokoff}, E., {Stark}, A.~A., {Story}, K., {Suzuki}, A.,
  {Thompson}, K.~L., {Tucker}, C., {Vanderlinde}, K., {Vieira}, J.~D.,
  {Vikhlinin}, A., {Wang}, G., {Yefremenko}, V., and {Yoon}, K.~W., ``{SPT-3G:
  a next-generation cosmic microwave background polarization experiment on the
  South Pole telescope},'' in [{\em Millimeter, Submillimeter, and Far-Infrared
  Detectors and Instrumentation for Astronomy
  VII}{\nolinebreak\hspace{0.1em}]},  {Holland}, W.~S. and {Zmuidzinas}, J.,
  eds., {\em Society of Photo-Optical Instrumentation Engineers (SPIE)
  Conference Series} {\bf 9153},  91531P (July 2014).

\bibitem{Fowler:07}
Fowler, J.~W., Niemack, M.~D., Dicker, S.~R., Aboobaker, A.~M., Ade, P. A.~R.,
  Battistelli, E.~S., Devlin, M.~J., Fisher, R.~P., Halpern, M., Hargrave,
  P.~C., Hincks, A.~D., Kaul, M., Klein, J., Lau, J.~M., Limon, M., Marriage,
  T.~A., Mauskopf, P.~D., Page, L., Staggs, S.~T., Swetz, D.~S., Switzer,
  E.~R., Thornton, R.~J., and Tucker, C.~E., ``Optical design of the atacama
  cosmology telescope and the millimeter bolometric array camera,'' {\em Appl.
  Opt.}~{\bf 46},  3444--3454 (Jun 2007).

\bibitem{Padin:08}
Padin, S., Staniszewski, Z., Keisler, R., Joy, M., Stark, A.~A., Ade, P. A.~R.,
  Aird, K.~A., Benson, B.~A., Bleem, L.~E., Carlstrom, J.~E., Chang, C.~L.,
  Crawford, T.~M., Crites, A.~T., Dobbs, M.~A., Halverson, N.~W., Heimsath, S.,
  Hills, R.~E., Holzapfel, W.~L., Lawrie, C., Lee, A.~T., Leitch, E.~M., Leong,
  J., Lu, W., Lueker, M., McMahon, J.~J., Meyer, S.~S., Mohr, J.~J., Montroy,
  T.~E., Plagge, T., Pryke, C., Ruhl, J.~E., Schaffer, K.~K., Shirokoff, E.,
  Spieler, H.~G., and Vieira, J.~D., ``South pole telescope optics,'' {\em
  Appl. Opt.}~{\bf 47},  4418--4428 (Aug 2008).

\bibitem{decadal}
Abazajian, K., Addison, G., Adshead, P., Ahmed, Z., Allen, S.~W., Alonso, D.,
  Alvarez, M., Amin, M.~A., Anderson, A., Arnold, K.~S., Baccigalupi, C.,
  Bailey, K., Barkats, D., Barron, D., Barry, P.~S., Bartlett, J.~G., Thakur,
  R.~B., Battaglia, N., Baxter, E., Bean, R., Bebek, C., Bender, A.~N., Benson,
  B.~A., Berger, E., Bhimani, S., Bischoff, C.~A., Bleem, L., Bock, J.~J.,
  Bocquet, S., Boddy, K., Bonato, M., Bond, J.~R., Borrill, J., Bouchet, F.~R.,
  Brown, M.~L., Bryan, S., Burkhart, B., Buza, V., Byrum, K., Calabrese, E.,
  Calafut, V., Caldwell, R., Carlstrom, J.~E., Carron, J., Cecil, T.,
  Challinor, A., Chang, C.~L., Chinone, Y., Cho, H.-M.~S., Cooray, A.,
  Crawford, T.~M., Crites, A., Cukierman, A., Cyr-Racine, F.-Y., de~Haan, T.,
  de~Zotti, G., Delabrouille, J., Demarteau, M., Devlin, M., Di~Valentino, E.,
  Dobbs, M., Duff, S., Duivenvoorden, A., Dvorkin, C., Edwards, W., Eimer, J.,
  Errard, J., Essinger-Hileman, T., Fabbian, G., Feng, C., Ferraro, S.,
  Filippini, J.~P., Flauger, R., Flaugher, B., Fraisse, A.~A., Frolov, A.,
  Galitzki, N., Galli, S., Ganga, K., Gerbino, M., Gilchriese, M., Gluscevic,
  V., Green, D., Grin, D., Grohs, E., Gualtieri, R., Guarino, V., Gudmundsson,
  J.~E., Habib, S., Haller, G., Halpern, M., Halverson, N.~W., Hanany, S.,
  Harrington, K., Hasegawa, M., Hasselfield, M., Hazumi, M., Heitmann, K.,
  Henderson, S., Henning, J.~W., Hill, J.~C., Hlozek, R., Holder, G.,
  Holzapfel, W., Hubmayr, J., Huffenberger, K.~M., Huffer, M., Hui, H., Irwin,
  K., Johnson, B.~R., Johnstone, D., Jones, W.~C., Karkare, K., Katayama, N.,
  Kerby, J., Kernovsky, S., Keskitalo, R., Kisner, T., Knox, L., Kosowsky, A.,
  Kovac, J., Kovetz, E.~D., Kuhlmann, S., Kuo, C.-l., Kurita, N., Kusaka, A.,
  Lahteenmaki, A., Lawrence, C.~R., Lee, A.~T., Lewis, A., Li, D., Linder, E.,
  Loverde, M., Lowitz, A., Madhavacheril, M.~S., Mantz, A., Matsuda, F.,
  Mauskopf, P., McMahon, J., Meerburg, P.~D., Melin, J.-B., Meyers, J., Millea,
  M., Mohr, J., Moncelsi, L., Mroczkowski, T., Mukherjee, S., Münchmeyer, M.,
  Nagai, D., Nagy, J., Namikawa, T., Nati, F., Natoli, T., Negrello, M.,
  Newburgh, L., Niemack, M.~D., Nishino, H., Nordby, M., Novosad, V., O'Connor,
  P., Obied, G., Padin, S., Pandey, S., Partridge, B., Pierpaoli, E., Pogosian,
  L., Pryke, C., Puglisi, G., Racine, B., Raghunathan, S., Rahlin, A.,
  Rajagopalan, S., Raveri, M., Reichanadter, M., Reichardt, C.~L., Remazeilles,
  M., Rocha, G., Roe, N.~A., Roy, A., Ruhl, J., Salatino, M., Saliwanchik, B.,
  Schaan, E., Schillaci, A., Schmittfull, M.~M., Scott, D., Sehgal, N.,
  Shandera, S., Sheehy, C., Sherwin, B.~D., Shirokoff, E., Simon, S.~M.,
  Slosar, A., Somerville, R., Staggs, S.~T., Stark, A., Stompor, R., Story,
  K.~T., Stoughton, C., Suzuki, A., Tajima, O., Teply, G.~P., Thompson, K.,
  Timbie, P., Tomasi, M., Treu, J.~I., Tristram, M., Tucker, G., Umiltà, C.,
  van Engelen, A., Vieira, J.~D., Vieregg, A.~G., Vogelsberger, M., Wang, G.,
  Watson, S., White, M., Whitehorn, N., Wollack, E.~J., Wu, W. L.~K., Xu, Z.,
  Yasini, S., Yeck, J., Yoon, K.~W., Young, E., and Zonca, A., ``Cmb-s4 decadal
  survey apc white paper,'' (2019).

\bibitem{zmxmanual}
{Zemax LLC}, {\em {OpticStudio 22.2 User Manual}} (2022).

\bibitem{2022Gallardo}
{Gallardo}, P.~A. and {CMB-S4 Collaboration}, ``{Freeform three mirror
  anastigmatic large aperture telescope and receiver optics for CMB-S4},''
  Applied Optics (In prep.).

\bibitem{10.1117/12.2312985}
Galitzki, N., Ali, A., Arnold, K.~S., Ashton, P.~C., Austermann, J.~E.,
  Baccigalupi, C., Baildon, T., Barron, D., Beall, J.~A., Beckman, S., Bruno,
  S. M.~M., Bryan, S., Calisse, P.~G., Chesmore, G.~E., Chinone, Y., Choi,
  S.~K., Coppi, G., Crowley, K.~D., Crowley, K.~T., Cukierman, A., Devlin,
  M.~J., Dicker, S., Dober, B., Duff, S.~M., Dunkley, J., Fabbian, G.,
  Gallardo, P.~A., Gerbino, M., Goeckner-Wald, N., Golec, J.~E., Gudmundsson,
  J.~E., Healy, E.~E., Henderson, S., Hill, C.~A., Hilton, G.~C., Ho, S.-P.~P.,
  Howe, L.~A., Hubmayr, J., Jeong, O., Keating, B., Koopman, B.~J., Kiuchi, K.,
  Kusaka, A., Lashner, J., Lee, A.~T., Li, Y., Limon, M., Lungu, M., Matsuda,
  F., Mauskopf, P.~D., May, A.~J., McCallum, N., McMahon, J., Nati, F.,
  Niemack, M.~D., Orlowski-Scherer, J.~L., Parshley, S.~C., Piccirillo, L.,
  Rao, M.~S., Raum, C., Salatino, M., Seibert, J.~S., Sierra, C., Silva-Feaver,
  M., Simon, S.~M., Staggs, S.~T., Stevens, J.~R., Suzuki, A., Teply, G.,
  Thornton, R., Tsai, C., Ullom, J.~N., Vavagiakis, E.~M., Vissers, M.~R.,
  Westbrook, B., Wollack, E.~J., Xu, Z., and Zhu, N., ``{The Simons
  Observatory: instrument overview},'' in [{\em Millimeter, Submillimeter, and
  Far-Infrared Detectors and Instrumentation for Astronomy
  IX}{\nolinebreak\hspace{0.1em}]},  Zmuidzinas, J. and Gao, J.-R., eds.,  {\bf
  10708},  1 -- 13, International Society for Optics and Photonics, SPIE
  (2018).

\bibitem{10.1117/12.2312971}
Gallardo, P.~A., Gudmundsson, J., Koopman, B.~J., Matsuda, F.~T., Simon, S.~M.,
  Ali, A., Bryan, S., Chinone, Y., Coppi, G., Cothard, N., Devlin, M.~J.,
  Dicker, S., Fabbian, G., Galitzki, N., Hill, C.~A., Keating, B., Kusaka, A.,
  Lashner, J., Lee, A.~T., Limon, M., Mauskopf, P.~D., McMahon, J., Nati, F.,
  Niemack, M.~D., Orlowski-Scherer, J.~L., Parshley, S.~C., Puglisi, G.,
  Reichardt, C.~L., Salatino, M., Staggs, S., Suzuki, A., Vavagiakis, E.~M.,
  Wollack, E.~J., Xu, Z., and Zhu, N., ``{Systematic uncertainties in the
  Simons Observatory: optical effects and sensitivity considerations},'' in
  [{\em Millimeter, Submillimeter, and Far-Infrared Detectors and
  Instrumentation for Astronomy IX}{\nolinebreak\hspace{0.1em}]},  Zmuidzinas,
  J. and Gao, J.-R., eds.,  {\bf 10708},  658 -- 673, International Society for
  Optics and Photonics, SPIE (2018).

\bibitem{2018SPIE10708E..3XO}
{Orlowski-Scherer}, J.~L., {Zhu}, N., {Xu}, Z., {Ali}, A., {Arnold}, K.~S.,
  {Ashton}, P.~C., {Coppi}, G., {Devlin}, M., {Dicker}, S., {Galitzki}, N.,
  {Gallardo}, P.~A., {Keating}, B., {Lee}, A.~T., {Limon}, M., {Lungu}, M.,
  {May}, A., {McMahon}, J., {Niemack}, M.~D., {Piccirillo}, L., {Puglisi}, G.,
  {Salatino}, M., {Silva-Feaver}, M., {Simon}, S.~M., {Thornton}, R., and
  {Vavagiakis}, E.~M., ``{Simons Observatory large aperture receiver simulation
  overview},'' in [{\em Millimeter, Submillimeter, and Far-Infrared Detectors
  and Instrumentation for Astronomy IX}{\nolinebreak\hspace{0.1em}]},
  {Zmuidzinas}, J. and {Gao}, J.-R., eds., {\em Society of Photo-Optical
  Instrumentation Engineers (SPIE) Conference Series} {\bf 10708},  107083X
  (July 2018).

\bibitem{2018SPIE10700E..3ED}
{Dicker}, S.~R., {Gallardo}, P.~A., {Gudmundsson}, J.~E., {Mauskopf}, P.~D.,
  {Ali}, A., {Ashton}, P.~C., {Coppi}, G., {Devlin}, M.~J., {Galitzki}, N.,
  {Ho}, S.~P., {Hill}, C.~A., {Hubmayr}, J., {Keating}, B., {Lee}, A.~T.,
  {Limon}, M., {Matsuda}, F., {McMahon}, J., {Niemack}, M.~D.,
  {Orlowski-Scherer}, J.~L., {Piccirillo}, L., {Salatino}, M., {Simon}, S.~M.,
  {Staggs}, S.~T., {Thornton}, R., {Ullom}, J.~N., {Vavagiakis}, E.~M.,
  {Wollack}, E.~J., {Xu}, Z., and {Zhu}, N., ``{Cold optical design for the
  large aperture Simons' Observatory telescope},'' in [{\em Ground-based and
  Airborne Telescopes VII}{\nolinebreak\hspace{0.1em}]},  {Marshall}, H.~K. and
  {Spyromilio}, J., eds., {\em Society of Photo-Optical Instrumentation
  Engineers (SPIE) Conference Series} {\bf 10700},  107003E (July 2018).

\bibitem{korsch1991reflective}
Korsch, D.,  [{\em {Reflective optics}}{\nolinebreak\hspace{0.1em}]}, Academic
  Press (1991).

\end{thebibliography}
\bibliographystyle{spiebib} 

\end{document}